# A Type System for Data Privacy Compliance in Active Object Languages

Chinmayi Prabhu Baramashetru[a] , Paola Giannini[b] , Silvia Lizeth Tapia
Tarifa[a] , and Olaf Owe[a]

a   Department of Informatics, University of Oslo, Oslo, Norway
b   DISSTE, University of Eastern Piedmont, Piedmont, Italy

**Abstract**   Data protection laws such as GDPR aim to give users unprecedented control over their personal data. Compliance with these regulations requires systematically considering information flow and interactions among entities handling sensitive data. Privacy-by-design principles advocate embedding data protection into system architectures as a default. However, translating these abstract principles into concrete, explicit methods remains a significant challenge. This paper addresses this gap by proposing a language-based approach to privacy integration, combining static and runtime techniques. By employing type checking and type inference in an active object language, the framework enables the tracking of authorised data flows and the automatic generation of constraints checked at runtime based on user consent. This ensures that personal data is processed in compliance with GDPR constraints. The key contribution of this work is a type system that gather the compliance checks and the changes to users consent and integrates data privacy compliance verification into system execution. The paper demonstrates the feasibility of this approach through a soundness proof and several examples, illustrating how the proposed language addresses common GDPR requirements, such as user consent, purpose limitation, and data subject rights. This work advances the state of the art in privacy-aware system design by offering a systematic and automated method for integrating GDPR compliance into programming languages. This capability has implications for building trustworthy systems in domains such as healthcare or finance, where data privacy is crucial.

## The Art, Science, and Engineering of Programming

## 1 Introduction

The General Data Protection Regulation (GDPR) [13] mandates transparent data processing and handling of personal data according to users' consent. In practice, such consent is often implicit, with organisations collecting data through complex privacy policies, leading to many cases of unlawful processing and data breaches. Techniques for GDPR enforcement are currently debated [31, 37], due to the regulation's extensive and sometimes ambiguous terminology. While GDPR emphasises privacy by design and default [11, 25] in Article 25, placing the responsibility on Data Controllers to implement appropriate technical measures, organisations often struggle to adapt their models and software for data privacy compliance.

Integrating data privacy principles into the implementation of systems is a long-standing challenge [16, 41]. While there have been efforts to develop privacy by design [37], many aspects of data privacy remain unexplored [22, 47]. Conventional programming languages lack robust support for privacy-specific properties, highlighting the need for data privacy compliance guarantees, especially in distributed settings, where multiple parties handle personal data. In previous work, Baramashetru et al. developed a Privacy-Aware Active Object language (P-AOL) [5], a compact imperative language based on the active object paradigm [9]. P-AOL supports features like concurrent objects, asynchronous communication, and dynamic object generation, it also integrates GDPR privacy notions, offering syntax for expressing data privacy policies for consent checking. The language models entities, users, purposes, consent, and restricted access to sensitive personal data, allowing users to manage their data privacy consent via special constructs. P-AOL's semantics ensure that personal data is accessed in line with users' consent by checking compliance at every execution point. However, such checks increase system complexity and slowing down execution since repeated checks are done over the same data values at different execution points in the same method call.

*Our Approach.* In this paper, we propose TyPAOL, a type-based extension of P-AOL a language with a type checking and constraint inference system that reduces runtime checking by performing static analysis of privacy constraints. TyPAOL uses program analysis to extract consent and privacy dependencies statically. By analysing how processes interact with user consent, TyPAOL tracks upfront which consent checks or modifications are done. Instead of enforcing privacy purely at runtime, TyPAOL collects constraints statically, allowing to reduce privacy compliance checks to scheduling points, e.g., method binding and activation, rather then at every execution point. While static typing ensures the absence of type errors and efficient code generation [29, 30], not all program properties can be enforced statically. Figure 1 shows our consent management framework, combining static (TyPAOL type system) and dynamic (typed operational

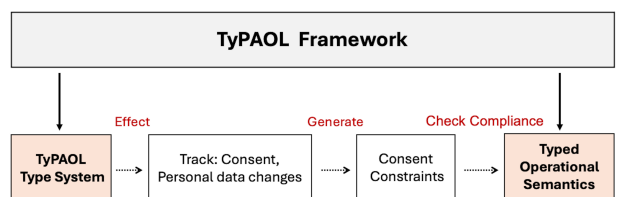

■ **Figure 1** Data privacy compliance guarantees with static and runtime checking.

semantics) enforcement. Dotted arrows indicate information flow across compilation and execution.The checking is done before runtime, as usual, but the runtime rules for method binding and method activation are modified, taking into account the constraints generated by the type system. At runtime, the static checks are already done, and only the mentioned constraints at method binding and method activation need to be checked. The approach guarantees that a program cannot stop in the middle of a method execution (process) due to lack of users' consent, since all checks are either already done statically or at the beginning of each process.

We adopt techniques from type systems, such as effects tracking [27] and constraint inference, to generate runtime checks [51]. Fields and variables are decorated with types, which also indicates whether their values may contain personal data or not. Actions on personal data require user consent, which must be computed at runtime since authorisations may vary with each method execution. Our type system *infers consent constraints*, which are checked at scheduling points. Producing such constraints requires tracing the effects of the statements changing variables and fields that hold personal data, and also tracing the effect of changes in the users' consent. We define a typed operational semantics, that checks the inferred consent constraints, we also show that the type checking and inference system is sound for this operational semantics.

Unlike pure runtime compliance enforcement in P-AOL, TyPAOL statically infers privacy constraints with reduced runtime checks while ensuring data privacy compliance. When developing privacy-aware type systems, managing aliasing and concurrent access to consent information is critical, TyPAOL enforces typing rules that prevent intra and inter-method aliasing, and introduces an interference check at method activation to ensure that processes modifying or depending on consent cannot interfere with each other. This guarantees that privacy constraints remain consistent across concurrent executions. While traditional security models rely on label-based information flow [12, 26, 52], TyPAOL enforces purpose-based restrictions that ensure data is processed only for user-approved purposes. Unlike conventional decentralised information flow control systems that dynamically enforce security labels [21, 53], TyPAOL integrates explicit purpose-based access control directly into its type system, aligning with GDPR's purpose limitation principle, along with other key GDPR principles. Existing static approaches such as Jif [26] and FlowCaml [40] cannot handle changing user preferences, whereas TyPAOL allows consent modifications at runtime while ensuring preemptive consent validation. In distributed settings, tracking data usage across multiple entities is complex, TyPAOL enables cross-entity privacy enforcement, ensuring compliance even when data flows between different entities.

In summary, we make the following contributions: 1) we propose a hybrid privacy compliance framework that integrates static type checking with runtime validation, offering optimised checks, 2) a detailed type-checking and constraint inference system, which collects consent-checking constraints, and 3) we prove the soundness of the inference method. To the best of our knowledge, TyPAOL is the first hybrid methodology integrating static type-based privacy constraints with runtime compliance enforcement in a systematic privacy-by-design approach.

*Paper outline.* Section 2 details P-AOL and GDPR integration with several examples. Section 3 emphasises on the motivation and key difference from P-AOL and introduces the *type checking and constraint inference system* of TyPAOL. Section 4 shows 1) how the type checking and inference system can be used at runtime in the form of *constraints* to optimise the privacy-aware checks of the language, and 2) the soundness of the type system. Section 5 discusses about system integration considerations for TyPAOL. Section 6 discusses the differences between P-AOL and TyPAOL along with related work and Section 7 concludes the paper. The complete operational semantics of the language, additional typing rules that are not included in the paper, and the soundness proof of the inference method are included in Appendix A, B, and C respectively.

## 2  Overview of P-AOL and the GDPR Integration

The Privacy-Aware Active Object Language (P-AOL) is a programming language that is explicitly designed to incorporate GDPR principles into its syntax and semantics. By adding privacy constructs directly into the language, P-AOL enables the systematic enforcement of key GDPR principles such as Conditions for Consent (Art. 7), Lawfulness, Fairness, and Transparency (Art. 5 Section. 1a), Accountability (Art. 5(2)), Privacy by design and default (Art. 25), Purpose limitation (Art. 5 Section. 1b), and Data subject rights (Art. 12-23), ensuring compliance of personal data handling. This section revisits P-AOL's integration of GDPR requirements and shows its functionality through examples.

### 2.1  GDPR Key Principles Addressed by P-AOL

The General Data Protection Regulation (GDPR) outlines strict rules for personal data processing, emphasising informed consent , purpose limitation, and data subject rights. P-AOL translates these principles into actionable language constructs, such as 1) *Consent Management (Art. 7)*: P-AOL allows users to dynamically grant or revoke consent. 2) *Lawfulness, Fairness, and Transparency (Art. 5 Section. 1a)*: P-AOL consent operations make processing activities visible to the user (transparency) and ensure that data handling is not only lawful but also aligned with the user's expectations of fairness. 3) *Purpose Limitation (Art. 5 Section. 1b)*: It ensures that data is processed only for specific, user-approved purposes. 4) *Entity-User Interaction:* P-AOL requires that entities (representing data controllers and data processors) must comply with the declared user consent when interacting with their personal data. 5) *Privacy-by-design (Art. 25):* Is a core principle of GDPR, and it is inherently supported by P-AOL. The language embeds privacy constructs directly into its operational semantics, ensuring that privacy policies are integral to the system behavior by default. 6) *Accountability (Art. 5(2)):* Through its delegation feature P-AOL ensures that each stakeholder's obligations are well-defined. The party retaining accountability (controller) has the legal responsibility to confirm valid user consent for all operations performed on personal data.

**Privacy Constructs in P-AOL** P-AOL introduces essential constructs for managing entities, purposes, and personal data. These constructs ensure privacy-aware programming and facilitate compliance with these key GDPR principles and enhances lawful personal data handling.

**entities** {HospEnt, GpEnt, PsEnt}; **purposes** {Healthcare, Emergency, Psychotherapy};
U alice := **new user**; $D$ patientRecord = "1234" **tag** ⟨{alice}, {Healthcare, Psychotherapy}⟩

■ **Figure 2** Example code showcasing privacy constructs in P-AOL.

- *Entities* are declared in a program and represent *data controllers (DCs)* and *data processors (DPs)* that handle personal and non-personal data (e.g., Hospital). Each object is associated with a declared entity, which can act as DC or DP. Objects can delegate the handling of personal data according to their own entity rights, while maintaining accountability for personal data handling. Delegation becomes effective if an object is allowed to temporarily take the access rights of the entity of the caller. This follows the GDPR principle of transfer of rights.

- *Purposes* are declared in the program and are reasons for handling personal data, giving a program the contexts in which personal data handling is allowed (e.g., Healthcare).

- *Data subjects (DSs)* are *users* whose personal data is being handled. P-AOL includes a special kind of objects that represent users so that it is possible to express how a service behaves when interacting with the users' personal data and consent.

- *Personal data* in a program are data values that are tagged, i.e., with user-specific ids and purposes. The tags allow to apply the right compliance checks when accessing personal data. P-AOL is designed to allow free handling of non-personal data while restricting personal data handling.

Consider the snippet of a P-AOL program in Figure 2, where we have entity declarations, HospEnt (for a hospital), GpEnt (for general practitioner) and PsEnt (for psychotherapist), and purposes declaration, followed by the creation of the user alice, the variable patientRecord is assigned with a personal data value that can be processed for Healthcare and Psychotherapy purposes and specifically for the user alice, here $D$ represent the type of the variable.

### 2.2 Overview of Operational Semantics of P-AOL

The runtime semantics of P-AOL centres around enforcing privacy compliance in alignment with GDPR principles. Every action performed on personal data during execution is dynamically checked to ensure it adheres to user consent and defined purposes. This section discusses these semantics informally, using various key concepts, that are embedded in the runtime semantics. The full list of the operational semantics rules for expressions, statements and sequences of statements are given in Appendix A.

**Key Concept of Privacy in the Runtime Semantics of P-AOL**    At runtime, P-AOL ensures GDPR compliance through the following:

- **Dynamic Consent Checks:** Before any data is processed, P-AOL verifies that the operation aligns with the users' current consent settings. If consent has been revoked or is insufficient, the operation is denied.

$$comply_U(\Sigma, \eta_{acc}, t) \quad = \quad \textbf{use} \in \mathscr{A}(\Sigma, \eta_{acc}, t)$$

$$comply_C(\Sigma, \eta_{acc}, t) \quad = \quad comply_U(\Sigma, \eta_{acc}, t) \wedge \textbf{collect} \in \mathscr{A}(\Sigma, \eta_{acc}, t)$$

$$comply_T(\Sigma, \eta_{acc}, t) \quad = \quad comply_U(\Sigma, \eta_{acc}, t) \wedge \textbf{transfer} \in \mathscr{A}(\Sigma, \eta_{acc}, t)$$

$$comply_S(\Sigma, \eta_{acc}, \eta_{callee}, t) \quad = \quad comply_C(\Sigma, \eta_{acc}, t) \wedge \textbf{store} \in \mathscr{A}(\Sigma, \eta_{callee}, t)$$

■ **Figure 3** Comply predicates to check users' consent in P-AOL.

- **Compliance Validation:** Operations on personal data are only allowed if there is compliance according to both the users' consent and the associated data tags. Each operation in the configuration checks this and hence compliance is validated.
- **Multi-owner Support:** P-AOL supports compliance checks for data with multiple owners and purposes.

### 2.2.1 Consent and Compliance Management

In P-AOL, consent gives entities the access rights to handle personal data and it is dynamic since it can be changed by users at any time via adding and removing privacy policies that specify which entities may handle personal data, for what purposes, and what actions they are allowed to perform, following GDPR terminology such as collect, transfer, use, store. Intuitively, P-AOL has the following interpretation for actions: *use* gives the right of read-access to personal data, *collect* for collecting personal data and is required for assigning them to local variables or receiving them as actual parameters of the methods, *store* for assigning personal data to fields in objects (i.e., assigning personal data to variables that live longer than the currently executing method), and *transfer* for transferring personal data to other objects. These specific language constructs align with GDPR's requirements for informed consent, ensuring compliance through dynamic updates.

Given the set of users $\widehat{u}$ and the set of purposes $\widehat{p}$ in P-AOL, personal data tagged with $\langle \widehat{u}, \widehat{p} \rangle$ must only be used for actions and purposes explicitly consented by all the users in $\widehat{u}$. To handle such data according to the consent of the users, we define the runtime predicates of Figure 3, where P-AOL's operational semantics rules, defined through these predicates, systematically enforce compliance during data handling. Given the users' consent $\Sigma$, the accountable entity $\eta_{acc}$ and the tag $t$, the predicate $comply_U$ is intuitively checked whenever an expression involving personal data is evaluated, ensuring that the user has explicitly granted the accountable entity $\eta_{acc}$ permission to *use* such data. here, the function $\mathscr{A}$ returns a set of actions that an entity $\eta$ can perform according to a consent $\Sigma$ and a tag $t$. If the result of this expression is assigned to a local variable, the system further verifies the users' consent to *collect* the data, as checked by the $comply_C$ predicate (e.g., local assignment operation). Similarly, if the data is passed as an argument to a method, the $comply_T$ predicate ensures the users have authorised the right to *transfer* the data (e.g., asynchronous call to a method operation). Finally, when the result expression is stored in a field of the current object (acting as the entity $\eta_{callee}$), the $comply_S$ predicate checks that the user has provided consent not just to use and collect the data to the accountable entity but also to *store* it in a field available to the entity $\eta_{callee}$ (e.g., field assignment operation). This layered approach ensures granular compliance with users' consent for every operation in a program involving personal data.

```
class Doctor(lb Lab){ ...
    handlePatientData(D patientData, U alice) {D localData;
        alice.addCon(({GpEnt, HospEnt}, { use, collect, store, transfer}, {Healthcare}));
        // assigns personal data to a local variable
        localData = patientData tag ⟨{alice}, {Healthcare}⟩; // check comply_C
        // ... process the data locally ...
        // async call: transfers the data to an external analytics entity
        Fut< D> f1 = lb!transferToAnalytics(localData); // checks comply_T }}
```

■ **Figure 4** Example drafted in P-AOL showcasing consent and comply checks.

```
class Therapist {
    D getPatientData( U userId) {
        // tag patient data for MentalHealthCounseling purpose
        D patientRecord1 = ("SessionNotes: Patient feels anxious") tag ⟨{userId}, {MentalHealthCounseling}⟩;
        // tag patient data for Research purpose
        D patientRecord2 = ("SessionNotes: Patient feels anxious") tag ⟨{userId}, {Research}⟩;
        //this fails due to purpose mismatch }}
```

■ **Figure 5** Example drafted in P-AOL showcasing purpose based data handling.

To showcase consent and compliance management, consider a hospital system where patients e.g., alice and bob, interact with entities such as HospEnt and GpEnt. In Figure 4, we can see a class Doctor with a method handlePatientData. By using the construct **addCon** inside the method body, user alice gives her consent. Afterwards, the variable patientData is tagged and evaluated, this operation needs to check permission to *use* the data, and, since it is assigned to the variable localData, it needs to check also permission to *collect*. These checks are expressed by the $comply_C$ predicate. Similarly, for an asynchronous call to *transfer* the personal data in the variable localData, the system along with *use* and *collect*, it checks also *transfer* rights. These checks are expressed by the $comply_T$ predicate.

### 2.2.2 Purpose-Based Data Processing

Given a scenario in which alice consents entity therapist to access her personal data for the purpose of MentalHealthCounseling using

$$\text{alice.} \textbf{addCon}((\{\text{PsEnt}\}, \{\textbf{use, collect, store}\}, \{\text{MentalHealthCounseling}\})).$$

In the code of Figure 5, when an instance of the class Therapist (that is associated to the entity PsEnt) tries to process the data for any other purpose (e.g., Research) using tags, the system detects a purpose mismatch with consent and therefore further processing in the method stops. Purpose-based processing is a cornerstone of GDPR compliance and is critical in P-AOL for ensuring that personal data is only used in ways explicitly authorised by the data subjects. This becomes even more important when data has multiple owners, as it ensures that the system respects the purposes approved by each owner individually, preventing conflicts or unintended use of personal data that is owned by various users, some being more or less permissive than others in their consent.

### 2.2.3 Dynamic Compliance Enforcement
**Unauthorised Access in a Hospital Information System**   Consider the hospital information system (HIS) scenario discussed earlier, where general practitioners (GP) interact

```
entities {HospEnt, GpEnt, LabEnt, DbEnt}; purposes {Diagnosis, Tests};
class GP(lb Lab , db DB) {
    // collect data and send it to Lab
    D diagnosisPatient(U u₁) {
        D pd = ("Alice", "123-45-6789") tag ({u₁}, {Diagnosis}) // authorized action to collect
        Fut< D > f₁ = lb!receiveData(pd); //send data to lab for tests}}

class Lab { D d₂;
    // receive data for processing
    D receiveData( D d₁ ) {
        d₂ = d₁ + "Test Results" ; // personal data combined with non personal
        Fut< D > f₂ = db!transferToDB( d₂); // violation of consent }}

class DB { D dataStore ;
    Unit AppendData() {
        dataStore = d₃; // does not have any rights }}

{ // MAIN BLOCK
    DB db := new DB( ) of DbEnt; GP gp := new GP(lab, db) of GpEnt;
    Lab lb := new Lab( ) of LabEnt;
    U alice := new user; U bob := new user;
    alice.addCon(({GpEnt}, {collect, use, transfer}, {Diagnosis}));
    alice.addCon(({LabEnt}, {use, collect, store}, {Tests}));
    gp!diagnosisPatient(alice); }
```

**■ Figure 6** A hospital information system (HIS) example, drafted in P-AOL showcasing prevention of unauthorised data handling.

with a centralised database (DB), and a laboratory system(Lab) to access patient information. We revisit this setup with a concrete operational flow, focusing on two key operational rules with names: *Assign1* (assigning values to local variables) and *Async-Call* (asynchronous method invocation). The following code highlights how P-AOL prevents unauthorised data handling using data privacy checks enforced by the operational semantics.

**System Setup**   The scenario described in the main block of Figure 6 establishes the initial state with entities such as GpEnt , LabEnt, DbEnt and user alice. Where alice grants GpEnt to **use**, **collect** and **transfer** her personal data for the purpose of Diagnosis, allows the LabEnt to **use**, **collect** and **store** her personal data for the purpose of Tests, but excludes the **transfer** actions on her personal data.

**Program Execution**   Figure 7 shows the flow of operations of the example. The object gp associated with GpEnt collects alice's data, tags it and does the assignment, these operations succeed due to the current valid consent. Next, the gp asynchronously transfers the data to the lab with valid consent. The lab receives the data and stores it locally, and tries to transfer the personal data to the database db asynchronously. In this scenario, gp.DiagnosisPatient() and lab.receiveData() execute as consent validates the use, collect and transfer actions. Whereas, db.transferToDB(db) fails since lab does not have transfer rights and execution cannot go on. However, if at a later point the

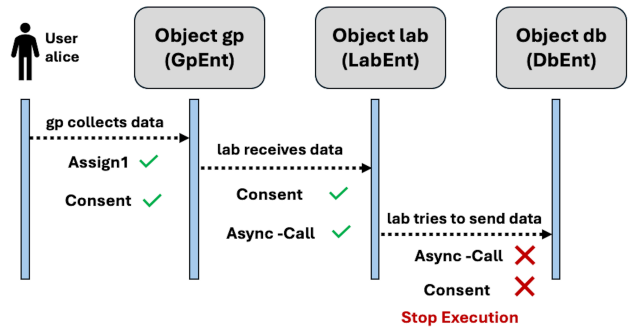

■ **Figure 7**  Consent flow for the hospital information system.

consent for transferring the data was added the method call could be executed. This example highlights how P-AOL enforces compliance by dynamically validating consent at each step. Unauthorised data handling operations are not executed, ensuring adherence to users' consent. By combining runtime validation with robust operational semantics, P-AOL allows the implementation of GDPR principles like *purpose limitation* and *lawful processing*.

### 2.2.4 Demonstrating Dynamic Consent Modification and Its Impact on Execution

**Example 1.** To show the flexibility and rigor of P-AOL s operational semantics, next example is an extended version of hospital information system example, which introduces dynamic consent modifications. We demonstrate how changes in users' consent during program execution affect the system's behavior. *Scenario*: In the main block, bob consents to his health personal data being used, collected and stored for Healthcare, Emergency and Psychotherapy. However, in another method, bob withdraws consent for transfer rights for Psychotherapy. This scenario explores the impact of this withdrawal on further processing. This example also gives a full length program written in P-AOL, making use of several of its constructs.

**A Hospital Information System**    In the hospital information system (HIS), drafted in P-AOL, and shown in Figure 8, doctors register new patients in the HIS. The HIS will communicate with a database DB, which stores the patients' personal data. Similar to our previous examples, we have highlighted in blue how personal data flows along the system. When a doctor registers a new patient via the method docRegisterPatient, which further calls the method registerPatient in the HIS, where the patient's data gets connected to an id and purposes via the **tag** construct, from there on, untagged data values become personal data in the system. The method registerPatient further calls in DB the method setPersonalData. When a doctor has an appointment with a call to method startApmnt, she calls the method requestPatientInfo in the HIS, which will further call the method requestPersonalData in the DB. Here the respective asynchronous calls will place the return value from the method call into the corresponding futures. The example takes advantage of the explicit control for synchronisation and data retrieval from a future (mailbox) [8] via the statement **get**.

In this scenario, bob revokes his consent, via the construct **remCon**, to PsEnt for the transfer action for the purpose of Psychotherapy. This dynamically updates the

```
entities {HospEnt, GpEnt, PsEnt}; purposes {Healthcare, Emergency, Psychotherapy};

class DB() { ... Unit setPersonalData(U u₁, D d₁){...};
    D requestPersonalData(U u₂){...}; ... }

class HIS(DB db) { ...
    Unit registerPatient(U u₃, D d₂) {
        if /* check that patient is not yet registered */ ...
            db!setPersonalData(U u₃, d₂ tag ⟨{u₃}, {Healthcare, Psychotherapy}⟩); ...}
    D requestPatientInfo(U u₄) { D d₃;
        Fut <D> f₁ := db!requestPersonalData(u₄); d₃ := f₁.get; return d₃; }
    Bool checkPatientStatus(D d₄, D d₅)
{ ... /*check if patient is registered and psychoanalysis form is signed*/ ...} ...}

class Doctor(HIS his) { ...
    Unit startApmnt(U u₅){ D d₆; Fut<D> f₂ = his!requestPatientInfo(u₅); d₆ := f₂.get; ...;
        ... /* start appointment here */ ...}
    Unit docRegisterPatient(U u₆, D d₇) { his!registerPatient(u₆ , d₇ ); }

class Therapist(HIS his) { ...
    Unit startCoupleApmnt(U u₇, U u₈) { D d₈; D d₉; Bool patientsStatus;
        Fut<D> f₃:= his!requestPatientInfo(u₇); Fut <D> f₄:= his!requestPatientInfo(u₈);
        d₈ := f₃.get; d₉ := f₄.get;
        Fut<Bool> f₅ = his!checkPatientStatus(d₈, d₉ );  patientsStatus:= f₅.get;
        if Valid(patientsStatus) {
            // consent modification happens here
            u₇.remCon({PsEnt}, {transfer}, {Psychotherapy}));
            // attempt to transfer data after consent revocation
            Fut D f₆ = his!requestPersonalData(u₇); // fails due to revoked consent
        } else {/*deny session*/ } } ... }

{ // MAIN BLOCK
    DB db := new DB() of HospEnt; HIS his := new HIS(db) of HospEnt;
    Doctor gp := new Doctor(his) of GpEnt; Therapist ps := new Therapist(his) of PsEnt ;
    U alice := new user; U bob := new user;

    alice.addCon({HospEnt}, {store, transfer}, {Healthcare));
    alice.addCon({GpEnt, PsEnt}, {collect, use, transfer}, {Healthcare, Psychotherapy}));
    bob.addCon({HospEnt}, {store, transfer}, {Healthcare));
    bob.addCon({HospEnt}, {transfer}, {Emergency}));
    bob.addCon({GpEnt}, {collect, use, transfer}, {Healthcare, Emergency, Psychotherapy }));
    bob.addCon({PsEnt}, {collect, use, transfer}, {Psychotherapy}));

    // ... register users in the HIS ...
    gp!docRegisterPatient(alice, δₐ ) ; gp!docRegisterPatient(bob, δᵦ ); ...
    gp!startApmnt(bob); ps!startCoupleApmnt(alice,bob ); }
```

■ **Figure 8** A hospital information system (HIS) example, drafted in P-AOL. Here $D$ abstracts away the personal data, and $\delta_a$, $\delta_b$ denote records of untagged data values.

system's consent disallowing further transfer operations for bob's data. Despite the revocation, when an asynchronous call is made in the method requestPersonalData by the therapist object ps, the required consent for transfer is no longer valid, this operation fails, stopping the execution of the method.

$PR ::= E\ P\ \overline{CL}\ \{\overline{A\,x};\ sts\}$  
$E ::= \textbf{entities}\ \widehat{\eta};$  
$P ::= \textbf{purposes}\ \widehat{p};$  
$A ::= B \mid C \mid \mathsf{U}$  
$B ::= T \mid T^{\#}$  
$T ::= Int \mid Bool \mid Unit \mid T * \cdots * T \mid \cdots$  
$CL ::= \textbf{class}\ C(\overline{A\,x})\ \{\ \overline{M}\ \}$  
$M ::= B[:\Lambda]\ m(\overline{A\,x})\ \{\ \overline{A\,x};\ sts\ \}$  
$\Lambda ::= (\widehat{vr}, t)$  
$t ::= \langle \widehat{x}, \widehat{p} \rangle$

$sts ::= \epsilon \mid s; sts \mid \textbf{return}\ \alpha$  
$s ::= \textbf{skip} \mid x := rhs \mid \textbf{this}.x := \alpha$  
$\quad \mid \textbf{if}\ \alpha\ s\ \textbf{else}\ s \mid [\textbf{Fut}\langle B \rangle\ x :=]\ vr@m(\overline{\alpha})$  
$\quad \mid \{sts\} \mid x.\textbf{addCon}(\rho) \mid x.\textbf{remCon}(\rho)$  
$rhs ::= \alpha \mid \textbf{new}\ C(\overline{\alpha})\ \textbf{of}\ \eta \mid \textbf{new user} \mid x.\textbf{get}$  
$@ ::= !\ |\ !!$  
$\alpha ::= e \mid e\ \textbf{tag}\ t$  
$e ::= v \mid vr \mid (\overline{e}) \mid e.n \mid e\ op\ e$  
$vr ::= x \mid \textbf{this}.x \mid \textbf{this}$  
$\rho ::= (\widehat{\eta}, \widehat{a}, \widehat{p})$  
$a ::= \textbf{use} \mid \textbf{collect} \mid \textbf{store} \mid \textbf{transfer}$

■ **Figure 9** Syntax of TyPAOL. Terms $\overline{x}$, $\widehat{p}$, etc. denote possibly empty list, respectively sets of the syntactic categories. Square brackets [ ] denote optional elements.

### 2.3 Critical Reflections on P-AOL's Limitations

The examples presented in this section illustrate that P-AOL implements a consent-based personal data access flow. Its dynamic consent handling model showcases the use of policy driven systems, enabling real-time enforcement of users consent. However, this flexibility introduces significant runtime fragility as consent violations are detected during execution at any point of the execution of a method.

This leads to the following issues: 1) *runtime complexity:* with the current semantics, consent must be repeatedly validated during execution, which increases runtime checks and operational uncertainty, 2) *unpredictable behavior:* dynamic policy changes can interrupt operations unexpectedly making it less robust for critical applications such as healthcare. Both of these problems are caused by *lack of analysis of the program context,* where the system does not use any static information about the program elements such as consent and everything is deferred to runtime. This motivates the introduction to TyPAOL and its static analysis.

## 3 TyPAOL: Extending P-AOL with Static Guarantees

In this section, we introduce TyPAOL which builds upon P-AOL by adding key syntactic and semantic extensions to ensure stronger static guarantees. We have demonstrated P-AOL in the previous section and showcased various language features and GDPR principles, we retain the same GDPR features in TyPAOL.. The enhancements in TyPAOL include extended types, method annotations and the parametrisation of the compliance checks, enabling precise consent and information flow tracking at the type level. TyPAOL mainly addresses the limitations of P-AOL discussed in Section 2.3.

The syntax of TyPAOL is given in Figure 9. A program *PR* includes entities *E*, purposes *P*, classes $\overline{CL}$, and a main block containing declarations $\overline{A\,x}$ and statements *sts*, mostly used for scenario instantiation, e.g., instantiates objects, users, and initial invocation of methods. The types *A* of the language, consist of ground types *B*, types *C* to specify the types for the instances of classes, i.e., the name of a class, and the type U for instances of the special kind of objects that represent users. Types annotated with # (e.g., $T^{\#}$) represent personal or privacy-restricted types for variables and personal data values, enforcing stricter access constraints than standard types. In the syntax of

```
entities {HospEnt, GpEnt, PsEnt}; purposes {Healthcare, Emergency, Psychotherapy};
class DB() { ... Unit setPersonalData(U u₁, D# d₁){...};
    D# : (∅, ⟨{u₂}, {Healthcare, Psychotherapy}⟩) requestPersonalData(U u₂) {...}; ... }

class HIS(DB db) { ...
    Unit registerPatient(U u₃, D# d₂) {
        if /* check that patient is not yet registered */ ...
            db!setPersonalData(U u₃, d₂ tag ⟨{u₃}, {Healthcare, Psychotherapy}⟩); ...}
    D# : (∅, ⟨{u₄}, {Healthcare, Psychotherapy}⟩) requestPatientInfo(U u₄) { D# d₃;
        Fut <D#> f₁ := db!requestPersonalData(u₄); d₃ := f₁.get; return d₃; }
    Bool# : ({d₄, d₅}, .) checkPatientStatus(D# d₄, D# d₅ )
        { ... /*check if patient is registered and psychoanalysis form is signed*/ ...} ...}
```

■ **Figure 10** An extension of part of the example in Figure. 8, with syntactic tag annotations.

P-AOL, although there was the notion of privacy-restricted types in the language, P-AOL did not develop a type system and hence there was no enforcement of type-based restrictions on personal data access.

A class $CL$ has a name $C$, formal parameters $\overline{x}$ of types $\overline{A}$ providing the initial values for its fields which are implicitly defined by them, and methods $\overline{M}$, we assume a basic type system that disallow fields of type U. In TyPAOL, a method consists of a signature and a method body. The signature includes a methods name, a list of input parameters, a return type and an optional privacy annotation associated to the type of the return value. The method body has local variable declarations followed by statements. Statements, $s$ and $sts$ and expressions, $e$, are as in P-AOL, except for the asynchronous method call, that is described in the delegation and accountability paragraph.

As discussed in Section 2, tags associate data values with the id of the users that own the data and the purposes for which such data can be handled. A privacy annotation $\Lambda$ may have tag dependency on the parameters i.e., if the return value depends on the tags of the input parameters, the annotation will include a set of these parameters. For example, if a method processes data for alice and bob, the tag of the return value tag might reflect both users. The method can also declare specific tags for the return value, e.g., a method could specify that its output is only associated with an explicit tag. To understand the annotations, in Figure 10 we consider part of the example shown in Figure 8 but now with syntactic addition of annotations. Parameterised tag annotations on methods allow to dynamically adapt data owners and purposes in the return type. Here methods requestPersonalData and requestPatientInfo are annotated with the expected tag of the return value, concretely, the tag contains the expected id of the data owner $u_i$ and the purposes for which their personal data can be handled e.g., {Healthcare, Psychotherapy}. Similarly, parameterised tags for the return value can also depend on the tags in the parameters e.g., the method checkPatientStatus in HIS, depends on tags of the parameters $d_4$ and $d_5$. The return value of checkPatientStatus is used by Therapist in the method startCoupleApmnt, which is a session with two users. Observe that the return value of the method (e.g., the current consent for psychoanalysis of a couple) belongs to both users involved in the session.

**Delegation and Accountability in TyPAOL**    In TyPAOL we make a difference between single (!) and double (!!) exclamation mark when making an asynchronous method call, to aligns with GDPR's definitions of delegation between DCs and DPs. Consider the example shown in Figure 8 where a single exclamation mark (!) represents a scenario where an object acting as DC such as a gp, temporarily delegates rights to the object his, to perform specific actions on personal data while still keeping the accountability, as in the call his!requestPatientInfo($u_5$) in the Doctor class, here the object gp maintains accountability, this follows the GDPR's requirement that DPs act as hired agents under the instructions of the DC. Whereas a double exclamation mark (!!) reflects a transfer accountability to the callee, where the DP operates independently from the DC for the specific operation. For instance, we can change the call gp!!startApmnt(bob) with GpEnt entity in Figure 8, so the gp assumes full responsibility for processing Bob's data, acting as a DC. This differentiation ensures a clear framework for delegation and accountability.

### 3.1 Type Checking and Constraint Inference for TyPAOL

In addition to standard type checking for object-oriented languages, the extended type system collects information that can be used at runtime in the form of constraints that must be satisfied to avoid violating users' consent. Rather than checking consent compliance at each execution step (as done in the operational semantics of P-AOL given in Appendix A), these checks are optimised by only being performed at method binding and activation. In this section, we first introduce the types used in the type checking and constraint generation and define some functions and relations on the type annotations, then we present the relevant typing rules for expressions and statements. The remaining typing rules can be found in Appendix B.

**Extended Types of TyPAOL**    In Figure 11, we define the extended types $ET$. The difference between the extended types and the program-level ones is that the extended ones track how the associated values change during the execution of the program. In particular, the tag annotation in data types tracks the combination of the privacy tags associated with the subexpressions of the expression whose evaluation produced its value. Specifically, $\widehat{vr}$ in a tag annotation says that the evaluated expression has as subexpressions the variables $vr \in \widehat{vr}$, and the $t$ in a tag annotation says that one of its subexpressions was explicitly tagged by $t$. Variables declared with a type $T^{\#}$ will have a non-empty tag annotation $\Lambda$ in the extended type system, becoming $T^{\Lambda}$. For objects, the extended type tracks, in addition to the class, the entity associated to the object, where _ denotes the entity associated to a parameter of a method, which will be known when the method is called. For user types, the type $\mathsf{U}\langle R \rangle$ records in $R$ the privacy policy changes from an initial consent, that can be either $\emptyset$ when the user is created or the consent of the actual parameter of a method, denoted by _, as for objects types. $R$ represents a symbolic description of the user's consent state, defined statically and whose exact value is only instantiated according to the actual parameters at runtime when the method (that uses the user variable) is activated. This allows the system to accurately track the user's consent within a method body. The predicate *is-init*($RT$) says if a reference types has been initialised, namely if either $RT = \mathsf{U}\langle R \rangle$ for some $R$ or $RT = C\langle \eta \rangle$ or $RT = C\langle\_\rangle$ for some $C$ and $\eta$.

$$
\begin{aligned}
ET &::= DT \mid RT & \text{extended types} \\
DT &::= T^{\Lambda} & \text{data types} \\
\Lambda &::= (\widehat{vr}, t) \mid \epsilon & \text{extended tag annotations} \\
RT &::= UT \mid CT & \text{reference to user and object types} \\
CT &::= C \mid C\langle \eta \rangle \mid C\langle \_ \rangle & \text{extended object types} \\
UT &::= \mathsf{U}\langle R \rangle \mid \mathsf{U} & \text{extended user types} \\
R &::= R + \rho \mid R - \rho \mid \_ \mid \emptyset & \text{policy expressions} \\
\Gamma &::= \overline{vr} : \overline{DT} & \text{data type environment} \\
\Delta &::= vr : RT, \Delta \mid x : FT, \Delta \mid \epsilon & \text{objects, users and futures type environment} \\
FT &::= \mathbf{Fut}\langle DT \rangle & \text{future types}
\end{aligned}
$$

■ **Figure 11** Extended types and environments for TyPAOL.

**Tag Annotations**  The composition of tag annotations is defined as follows:

$$
\epsilon \bullet (\widehat{vr}, t) = (\widehat{vr}, t) \bullet \epsilon = (\widehat{vr}, t) \quad (\widehat{vr}, t) \bullet (\widehat{vr'}, t') = (\widehat{vr} \,\cup\, \widehat{vr'}, t \oplus t')
$$

It combines all the tags of the variables with the explicit tags by performing the operation $\oplus$. Combining with $\epsilon$ (an empty annotation) does not affect $\Lambda$. Tag annotations follows a *partial order* $\sqsubseteq$, defined as

$$
\text{for all } \Lambda \text{ , } \Lambda \sqsubseteq \epsilon \qquad (\widehat{vr}, t) \sqsubseteq (\widehat{vr'}, t') \quad \text{iff} \quad \widehat{vr'} \subseteq \widehat{vr} \,\wedge\, t \sqsubseteq t'
$$

The partial order $\Lambda \sqsubseteq \Lambda'$ holds if $\Lambda'$ is less restrictive than $\Lambda$. The annotation $\epsilon$, imposing no restrictions, is the maximum element in the partial order (for variables or fields that are expected to be associated with non-personal data values). The tag $(\emptyset, \langle \emptyset, \mathscr{P} \rangle)$, where $\mathscr{P}$ represents all program purposes, is the least restrictive tag for personal data. Thus, for any $\Lambda \neq \epsilon$ we have $\Lambda \sqsubseteq (\emptyset, \langle \emptyset, \mathscr{P} \rangle)$. As for tags, the composition of annotations produces more restrictive annotations, i.e., $\Lambda_1 \bullet \Lambda_2 \sqsubseteq \Lambda_i \quad$ for $1 \leq i \leq 2$ and $\Lambda_1 \bullet \Lambda_2$ is the greatest lower bound of $\Lambda_1$ and $\Lambda_2$.

We define a *substitution operator*, $S(\Lambda, vr, \Lambda')$, that replaces the variable $vr$ in $\Lambda$ with the tag annotation $\Lambda'$. If the variable is not in $\Lambda$ the annotation is unchanged otherwise we remove $vr$ from the set of variables in $\Lambda$ and compose $\Lambda$ and $\Lambda'$. That is

$$
S(\Lambda, vr, \Lambda') = \begin{cases} (\widehat{vr'}, t) \bullet \Lambda' & \text{if } \Lambda = (\{vr\} \cup \widehat{vr'}, t) \\ \Lambda & \text{otherwise} \end{cases}
$$

We now introduce most of the *typing rules for expressions and statements*. The missing ones can be found in Appendix B. A typing rule derives a typing judgment (located below the line), from a number of (possibly zero) typing judgments (located above the line), and may have on the side some conditions that have to hold.

**Typing Rules for Expressions**  Let $\Gamma$ be the *type environment* for variables associated with data, and $\Delta$ for variables associated with objects, users and futures, see Figure 11. The judgement $\Gamma; \Delta \vdash \alpha : ET$ holds if the expression $\alpha$ has type $ET$ in the type environments $\Gamma$ and $\Delta$.

Let us look at the two most significant rules:

$$
\frac{(\text{T-OP})}{\dfrac{\Gamma; \Delta \vdash e_i : T_i^{\Lambda_i} \quad i \in \{1,2\}}{\Gamma; \Delta \vdash e_1 \, op \, e_2 : T^{\Lambda_1 \bullet \Lambda_2}}} \; op : T_1 \times T_2 \to T
\qquad
\frac{(\text{T-E-TAG})}{\dfrac{\Gamma; \Delta \vdash e : T^{\Lambda}}{\Gamma; \Delta \vdash e \; \mathbf{tag} \; \langle \widehat{x}, \widehat{p} \rangle : T^{\Lambda \bullet (\emptyset, \langle \widehat{x}, \widehat{p} \rangle)}}} \; \begin{aligned} & x \in \widehat{x} \implies \\ & \Delta(x) {=} UT \wedge \textit{is-init}(UT) \end{aligned}
$$

$$\begin{aligned}
Cnstr_U(\Lambda, \Delta) &= \{A_{Cnstr}(\textbf{use}, \Lambda, \Delta)\} \\
Cnstr_C(\Lambda, \Delta) &= Cnstr_U(\Lambda, \Delta) \cup \{A_{Cnstr}(\textbf{collect}, \Lambda, \Delta)\} \\
Cnstr_T(\Lambda, \Delta) &= Cnstr_U(\Lambda, \Delta) \cup \{A_{Cnstr}(\textbf{transfer}, \Lambda, \Delta)\} \\
Cnstr_S(\Lambda, \Delta) &= Cnstr_C(\Lambda, \Delta) \cup \{A_{Cnstr}(\textbf{store}, \Lambda, \Delta)\}
\end{aligned}$$

■ **Figure 12** Constraint generation to check users' consent in TyPAOL.

In Rule (T-OP) we use the operator • to produce the tag annotation of the result of the expression. Note that the type of the expression will be annotated with a non-empty label expression (typed as personal data) if either one of its operands contains personal data. For a tagged expression, Rule (T-E-TAG) requires the expression to have a data type and the variables $\widehat{x}$ in the tag must be defined and have user types. We assume that $\widehat{p}$ is a subset of the purposes defined for the program. We define the annotation with an empty set of variables and the tag $\langle \widehat{x}, \widehat{p} \rangle$, which we then combine with the tag annotations $\Lambda$ of the expression $e$.

**Typing Rules for Statements**  The *type judgement for statements* $\Gamma; \Delta \vdash s \triangleright \Gamma'; \Delta' \mid \boldsymbol{Cn}$ and for *sequences of statements* $\Gamma; \Delta \vdash sts \triangleright DT \mid \Gamma'; \Delta' \mid \boldsymbol{Cn}$ combine the checking of compatibility of types (as done for expressions) with the inference of the constraints $\boldsymbol{Cn}$, which is required to be satisfied to not stop execution due to violation of privacy requirements. Such requirements will be checked at runtime in Rule (ACTIVATE) of the operational semantics. The type judgement tracks the effect of the statements via $\Gamma'$ and $\Delta'$, where $\Gamma'$ tracks the changes to the tag annotations of variable containing personal data types due to the assignments, and $\Delta'$ the adding/removing of policies to the users. For sequences of statements, *DT* is the return type of the expression in a **return**-statement. *DT* is of type *Unit* if we have a sequence of statements that does not end with a **return**-statement.

**Constraints and Constraint Generation**  Constraints arise from handling statements referring to personal data, i.e., containing expressions whose evaluation produces values that have a tag $t = \langle \widehat{u}, \widehat{p} \rangle$. To execute such statements according to the operational semantics, all the users in $\widehat{u}$ must have given explicit permission for one of the purposes in $\widehat{p}$ and for the actions required by the specific statement. In Figure 3 we show the comply predicates which states the actions needed for the operations in the language. The type system infers action constraints of the shape
$$A_{Cnstr}(act, \Lambda, \Delta) \,,$$
Tag annotation $\Lambda$ collects information for the tag. The environment $\Delta$ links users' variables to consent expressions, collecting changes in the consent. $\Lambda$ and $\Delta$ reflect the data privacy related effect from the start of the method body to the current statement being analysed. From these constraints, we can derive the runtime predicates of Figure 3 by instantiating $\Lambda$ and $\Delta$ according to the current state in the configuration. The type system generates constraints using the auxiliary predicates defined in Figure 12. Each predicate takes the tag annotation $\Lambda$, which if $\epsilon$ (empty), it generates no constraints, otherwise generates the corresponding constraints as defined in Figure 12.

The typing rules use the predicate *CmpTy* (compatibility of types), defined in Figure 13, that compares the types in the arguments and holds if a value of the type to

**A Type System for Data Privacy Compliance in Active Object Languages**

$CmpTy(A_1 \ldots A_n, ET_1 \ldots ET_n) = true$, if $CmpTy(A_i, ET_i')$ for all $i \in \{1, \ldots, n\}$
$CmpTy(C, CT) = true$, if $CT = C\langle \eta \rangle \ \lor \ CT = C\langle \_ \rangle$ $\qquad CmpTy(U, U\langle R \rangle) = true$
$CmpTy(T, T^\epsilon) = true$ $\quad CmpTy(T^\#, T^\Lambda) = true$
$CmpTy(A, ET) = false$, otherwise

■ **Figure 13** Definition of the predicate $CmpTy$ that compares the types in the arguments.

<span style="color:purple">(T-ASGN-FIELD)</span>

$$\frac{\Gamma; \Delta \vdash \alpha : T^{\Lambda'}}{\Gamma, \textbf{this}.x : T^\Lambda; \Delta \vdash \textbf{this}.x := \alpha \rhd \Gamma, \textbf{this}.x : T^{\Lambda[\Lambda']}; \Delta \mid Cnstr_S(\Lambda[\Lambda'], \Delta)} \quad \Lambda' = \epsilon \lor \Lambda \neq \epsilon$$

<span style="color:purple">(T-NEW-OBJ)</span>

$$\frac{\Gamma; \Delta \vdash \alpha_i : ET_i \ \ i \in \{1, \ldots, n\}}{\Gamma; \Delta, x : C \vdash x := \textbf{new } C(\overline{\alpha}) \textbf{ of } \eta \rhd \Gamma; \Delta, x : C\langle \eta \rangle \mid \emptyset} \quad \begin{array}{l} C \in PR \quad fields(C) = \overline{A}\,\overline{x} \quad CmpTy(\overline{A}, \overline{ET}) \\ \forall \ j \in \{1, \ldots, n\} \ A_j = DT \implies ET_j = T^\epsilon \end{array}$$

<span style="color:purple">(T-NEW-USER)</span>

$$\frac{}{\Gamma; \Delta, x : U \vdash x := \textbf{new user} \rhd \Gamma; \Delta, x : U\langle \emptyset \rangle \mid \emptyset}$$

<span style="color:purple">(T-READ-FUT)</span>

$$\frac{}{\Gamma, y : T^\Lambda; \Delta, x : \textbf{Fut}\langle T^{\Lambda'} \rangle \vdash y := x.\textbf{get} \rhd \Gamma, y : T^{\Lambda[\Lambda']}; \Delta, x : \textbf{Fut}\langle T^{\Lambda'} \rangle \mid Cnstr_C(\Lambda[\Lambda'], \Delta)} \quad \Lambda' = \epsilon \lor \Lambda \neq \epsilon$$

<span style="color:purple">(T-ASYN-CALL)</span>

$$\frac{\Gamma; \Delta \vdash \alpha_i : ET_i \ \ i \in \{1, \ldots, n\}}{\Gamma; \Delta, vr : CT \vdash \textbf{Fut}\langle B' \rangle \ x := vr@m(\overline{\alpha}) \rhd \Gamma; \Delta, vr : CT, x : \textbf{Fut}\langle T^{\Lambda'} \rangle \mid \bigcup_{j \in J} Cnstr_T(\Lambda_j, \Delta)}$$

where the side conditions are

$CT = C\langle \eta \rangle \ \lor \ C\langle \_ \rangle \quad B\ m(\overline{A}\,\overline{x}){:}\Lambda \ldots \textbf{OK in } C \quad CmpTy(B\overline{A}, B'\overline{ET}) \quad J = \{j \mid ET_j = T_j^{\Lambda_j} \text{ for some } T_j^{\Lambda_j}\}$
$\Lambda' = S(\Lambda, x_j, \Lambda_j)_{j \in J}[x_k/\alpha_k]_{k \in \{1, \ldots, n\} \setminus J} \quad \forall \ h \neq k \in \{1, \ldots, n\} \ (A_h = U \land A_k = U) \implies \alpha_h \neq \alpha_k$

<span style="color:purple">(T-SEQ)</span>

$$\frac{\Gamma; \Delta \vdash s \rhd \Gamma'; \Delta' \mid \textbf{Cn} \qquad \Gamma'; \Delta' \vdash sts \rhd DT \mid \Gamma''; \Delta'' \mid \textbf{Cn}'}{\Gamma; \Delta \vdash s; sts \rhd DT \mid \Gamma''; \Delta'' \mid \textbf{Cn} \cup \textbf{Cn}'}$$

<span style="color:purple">(T-RETURN)</span>

$$\frac{\Gamma; \Delta \vdash \alpha : T^\Lambda}{\Gamma; \Delta \vdash \textbf{return } \alpha \rhd T^\Lambda \mid \Gamma; \Delta \mid Cnstr_T(\Lambda, \Delta)}$$

<span style="color:purple">(T-ADD-CON)</span>

$$\frac{}{\Gamma; \Delta, x : U\langle R \rangle \vdash x.\textbf{addCon}((\widehat{\eta}, \widehat{a}, \widehat{p})) \rhd \Gamma; \Delta, x : U\langle R + \rho \rangle \mid \emptyset} \quad \widehat{\eta} \subseteq E \ \land \ \widehat{p} \subseteq P$$

<span style="color:purple">(T-REM-CON)</span>

$$\frac{}{\Gamma; \Delta, x : U\langle R \rangle \vdash x.\textbf{remCon}((\widehat{\eta}, \widehat{a}, \widehat{p})) \rhd \Gamma; \Delta, x : U\langle R - \rho \rangle \mid \emptyset} \quad \widehat{\eta} \subseteq E \ \land \ \widehat{p} \subseteq P$$

■ **Figure 14** Representative set of typing rules for single statements in TyPAOL.

the right can be assigned to a variable of the type to the left. Compatibility for $A$ and $ET$ types, including users and objects types, is as expected. This is also the case for list of these types. For compatibility of two tagged data types, $CmpTy(T^\Lambda, T^{\Lambda'})$ ensures that if $\Lambda = \epsilon$, then also $\Lambda' = \epsilon$, so we cannot assign a data value containing personal data to a field or variable that is expecting a data value with non-personal data.

Figure 14 contains a representative set of typing rules for single statements. The field assignment Rule (T-ASGN-FIELD), requires that the type of the expression to be assigned is compatible with the one of the variable. The type of the assigned variable in the resulting environment is modified to reflect the new tag annotation. Here

$$\Lambda[\Lambda'] = \begin{cases} \Lambda' & \text{if } \Lambda' \neq \epsilon \ \vee \ \Lambda = \epsilon \\ (\emptyset, \langle \emptyset, \mathscr{P} \rangle) & \text{otherwise} \end{cases}$$

So a variable that was declared to hold personal data does not become non-personal, but it gets the most permissive tag for personal values. The constraints generated reflect the fact that data are used and collected (by the accountable entity) and then stored in the current object. Rule (T-ASGN-OBJ) requires the class of the object $C$ to be defined in the program, and that the actual parameters passed to the constructor have a compatible types with the ones specified in the definition of the class. Moreover, there are no actual parameters $\alpha_i$ containing personal data, as asserted by the fact that all the data types $T_j$ have an empty label expression $\epsilon$. Note that the type $C$ means that this is a local variable which has not been initialised yet. Rule (T-NEW-USER) ensures that variables of type user are assigned only once, and only with a newly created user object. This restriction guarantees that no other variable or parameter in the method is an alias to the newly created user. The result of accessing a future with a **get** expression, Rule (T-READ-FUT), can be assigned to a variable, if the type of its content is compatible with the one of the variable. The constraints generated take into account the fact that data are used and collected (see Figure 12).

In the asynchronous calls, Rule (T-ASYN-CALL), a future variable is declared to hold the result of the method, the callee must be initialised, its class must contain the method declaration and the type of the expressions, actual parameters of the method must be compatible with the types of the formal parameters. Moreover, the result type must be compatible with the one declared for the future variable. For parameters referring to users, which must be variables, we check that they must be different variables, so that in the called method, there is no aliasing between two different formal parameters. This together with the restriction that to local variable of type user we can only assign a new user, prevents intra-method aliasing. For each parameter, we generate the constraint saying that data are used and transferred (see Figure 12). Note that, the method will be executed asynchronously in a (possibly) different object, so the the constraints of the method are not added to the one of the evaluation of the parameters. The future variable is added to the data environment of the continuation of the statement. The tag $\Lambda'$ of the future variable is obtained by the one of the result of the method $\Lambda$. Since $\Lambda$ may contain references to the formal parameters of $m$, we use the substitution defined for tag annotations for the formal parameters which have a data type and the standard substitution for parameters of user type.

Rule (T-SEQ) collects the changes of the typing environments $\Gamma$ and $\Delta$ for the different statements and collects the different generated constraints **Cn**. Rule (T-RETURN) produces the constraint saying that the result is used and transferred (see Figure 12). Rules (T-ADD-CON) and (T-REM-CON) modify the environment $\Delta$ by adding/removing policies to variables referring to users. The rules check that the entity and purposes are defined in the program.

(T-CLASS)

$$\frac{\forall M \in \overline{M} \ \exists \boldsymbol{Cn} \quad M, \boldsymbol{Cn} \ \textbf{OK in } C \qquad \forall A x \in \overline{A} \, \overline{x} \ A \neq U}{\textbf{class } C(\overline{A} \, \overline{x}) \ \{ \ \overline{M} \ \} \ \textbf{OK}}$$

(T-METH)

$$\frac{\Gamma_{lc} \ \Gamma_{p\&f}; \Delta_{lc} \ \Delta_{p\&f}, \textbf{this} : C \vdash sts \triangleright T^{\Lambda'} \mid \Gamma; \Delta \mid \boldsymbol{Cn}}{B \ m(\overline{A} \, \overline{x}) : \Lambda \{ \ \overline{A' \, y}; \ sts \}, \langle \boldsymbol{Cn}, \boldsymbol{Um} \rangle \ \textbf{OK in } C}$$

$fields(C) = \overline{A''} \, \overline{z}$
$B \ m(\overline{A} \, \overline{x}) : \Lambda \{ \ \overline{A' \, y}; \ sts \} \in C$
$TagOk(\Lambda, \overline{A} \, \overline{x})$
$CmpTy(B, T^{\Lambda})$ and $\Lambda \sqsubseteq \Lambda'$
$(\Gamma_{lc}, \Delta_{lc}) = TE_{lc}(\overline{A'} \, \overline{y})$
$(\Gamma_{p\&f}, \Delta_{p\&f}) = TE_{p\&f}(\overline{A} \, \overline{x} \, \overline{A''} \, \textbf{this}.\overline{z})$
$\boldsymbol{Um} = \{x | \Delta(x) = U\langle R \rangle \wedge R \neq \_ \wedge \ R \neq \emptyset \}$

■ **Figure 15** Well-typed method and class declaration in TyPAOL.

$TE_{lc}(A \, x, \overline{A} \, \overline{x}) =$
$\quad \textbf{let } (\Gamma_{lc}, \Delta_{lc}) = TE_{lc}(\overline{A} \, \overline{x}) \ \textbf{in case } A$

$\quad T \quad \implies (\Gamma_{lc}, x : T^{\epsilon}, \Delta_{lc})$
$\quad T^{\#} \quad \implies (\Gamma_{lc}, x : T^{(\emptyset, \langle \emptyset, \mathscr{P} \rangle)}, \Delta_{lc})$
$\quad C \quad \implies (\Gamma_{lc}, \Delta_{lc}, x : C)$
$\quad U \quad \implies (\Gamma_{lc}, \Delta_{lc}, x : U)$

$TE_{p\&f}(A \, vr, \overline{A} \, \overline{vr}) =$
$\quad \textbf{let } (\Gamma_{p\&f}, \Delta_{p\&f}) = TE_{p\&f}(\overline{A} \, \overline{vr}) \ \textbf{in case } A$

$\quad T \quad \implies (\Gamma_{p\&f}, vr : T^{\epsilon}, \Delta_{p\&f})$
$\quad T^{\#} \quad \implies (\Gamma_{p\&f}, vr : T^{(\{vr\}, \langle \emptyset, \mathscr{P} \rangle)}, \Delta_{p\&f})$
$\quad C \quad \implies (\Gamma_{p\&f}, \Delta_{p\&f}, vr : C\langle \_ \rangle)$
$\quad U \quad \implies (\Gamma_{p\&f}, \Delta_{p\&f}, vr : U\langle \_ \rangle)$

■ **Figure 16** Initial Environment Operators $TE_X$.

Note that in Rule (T-ASYNCALL) the constraint that two actual parameters cannot refer to the same user is important, since say $u$ is the actual parameter bound to both $x$ and $y$: if first $\rho$ is added to $x$ and then $\rho'$ is removed from $y$, after these two statements we would have in the environment $x : U\langle R + \rho \rangle, y : U\langle R - \rho \rangle$. Now in the type environment generated the order of these changes to consent is lost. For example, assume that $R = \_$, indicating that we start with the consent of the actual parameter, $\rho = \rho' = (\{\eta\}, \{a\}, \{p\})$ for some $\eta$, $a$ and $p$ and at the beginning of the method $\Sigma(u) = \emptyset$. In the consent environment generation of the instantiation of constraints, see Figure 20, the instantiation of $U\langle \_ + \rho \rangle$ followed from the one of $U\langle \_ - \rho \rangle$ would produce $\emptyset$, as expected, whereas if instantiation is done in the reverse order we obtain $[a \mapsto \{(\eta, p)\}]$, producing the wrong consent environment. In addition to the check Rule (T-ASYNCALL), to avoid aliasing we allow only new users in the right-hand-side of an assignment to a variable referring to a user.

The *type judgement for class declaration*, Rule (T-CLASS) of Figure 15, checks that all the methods of the class are well-typed and that no field are of user type. The *type judgement for method declaration*, Rule (T-METH) of Figure 15, checks that the body of the method is well-typed in the environments containing the initial extended types for fields, parameters and local variables, and the metavariable **this** associated to the type of the current class. The operators $TE_{lc}$ and $TE_{p\&f}$, see Figure 16, produce the data and object/user type environments, which associate parameters, variables and fields to their initial extended types. The operator $TE_{lc}$ considers the declaration of the local variables of the method: non-personal data are given tag annotation $\epsilon$ whereas for personal data the annotation specifies that the variable does not depend on parameters or fields annotations and the tag component of the annotation specifies the least restrictive tag for personal data. For users and objects, we assign as the initial extended type, the original type. The operator $TE_{p\&f}$ considers either the declarations of the parameters of the method or the fields of the class enclosing it: non-personal data are given tag annotation $\epsilon$ whereas for personal data we indicate that the variable depends on the corresponding variable with $vr$ and the tag component of an annotation is the least restrictive tag. The extended user type $\mathsf{U}\langle\_\rangle$ indicates that when the method starts, the user has the runtime consent and instances associated to the user are bound to the variable, which will be substituted for $\_$ when instantiating the constraints in Rule (Activate). For objects, we assign as initial extended type the original type. For the return type, we check that the tag annotation $\Lambda$ of the definition of the method contains only references to variables in the parameter list and that they are of the right type, this is done by the predicate $TagOk(\Lambda, \overline{A}\,\overline{x})$, which is defined by:

$$\Lambda = (\hat{y}, (\hat{z}, \hat{p})) \ \wedge \ \forall z \in \hat{z} \ \ \mathsf{U}\, z \in \overline{A}\,\overline{x} \ \wedge \ \forall \ y \in \hat{y} \ \exists \ T \ \ T\, y \in \overline{A}\,\overline{x}$$

Finally, the return type must have a tag annotation that is less restrictive than the one specified in the declaration of the method, and also the return type must be compatible with the declared one. The method definition is enriched with its set of constraints and the user variables that are modified by its execution. This information will be needed in the operational semantics Rules (Bind-Mtd) and (Activate).

A program is well-typed if all its classes and the main block are well-typed. Moreover the set of constraints generated for the main block is empty, i.e., there should not be uses of personal data.

**Definition 1.** *Let* $PR = E \ P \ \overline{CL} \ \{\overline{A}\,\overline{x};\ sts\}$, $PR$ *is* well-typed *if*

- $CL$ **OK** *for all* $CL \in \overline{CL}$ *and*
- $\Gamma_{lc}; \Delta_{lc} \vdash sts \vartriangleright Unit \,|\, \Gamma; \Delta \,|\, \emptyset$ *where* $(\Gamma_{lc}, \Delta_{lc}) = TE_{lc}(\overline{A}\,\overline{x})$.

The main block is used to establish the initial scenario of the program.

**Example 2.** Consider type checking of the method from our hospital information system example in Figure 8, where rpd is the short versions of requestPersonalData.

> $D^{\#}$ : $(\_, \langle\{u_4\}, \{\mathsf{Healthcare}, \mathsf{Psychotherapy}\}\rangle)$ requestPatientInfo($\mathsf{U}\ u_4$)
>
> $\{D^{\#} d_3;\ \mathsf{Fut}\langle D^{\#}\rangle\, f_1 := \mathbf{this}.\mathsf{db}!\mathsf{requestPersonalData}(u_4);\ d_3 := f_1.\mathbf{get};\ \mathbf{return}\ d_3;\ \}$

Let the initial environments for the typing of the body of the method, generated by the $TE_X$ operators, be

- $\Gamma_{lc} = d_3 : D^{(\emptyset,\langle\emptyset,\mathscr{P}\rangle)}$ and $\Gamma_{p\&f} = \emptyset$ and $\Gamma = \Gamma_{lc} \cup \Gamma_{p\&f}$.
- $\Delta_{lc} = \emptyset$ and $\Delta_{p\&f} = u_4 : \mathsf{U}\langle\_\rangle, \mathbf{this}.\mathsf{db} : \mathsf{DB}, \mathbf{this} : \mathsf{HIS}$ and $\Delta = \Delta_{lc} \cup \Delta_{p\&f}$

In Figure 17 we have the type derivations for the three statements of the function body (to save space we abbreviated the name of the rules and purposes).

- The first statement is an asynchronous method call to method requestPersonalData of the class DB (see $\mathscr{D}_1$). Rule (t-asyn-call) is applied, and the future $f_1$ is added to the environment $\Delta$. The type associated is the instance of tag annotation of the method in which $u_2$ (formal parameter) is replaced by the actual one $u_4$. Note also that in the syntax of the input code we can omit the set of variables of the

$$\mathcal{D}_1 \quad \dfrac{\dfrac{}{\Gamma;\Delta \vdash u_4 : U(\_)} \text{ T-A-I}}{\Gamma;\Delta \vdash \textbf{Fut}\langle D^{\#}\rangle \, f_1 := \textbf{this}.db!rpd(u_4) \rhd \Gamma;\Delta_1 \,|\, \emptyset} \text{ T-A-C} \quad \mathcal{D}_2 \quad \dfrac{}{\Gamma;\Delta_1 \vdash d_3 := f_1.\textbf{get} \rhd \Gamma_1;\Delta_1 \,|\, Cn_1} \text{ T-R-F}$$

$$\mathcal{D}_3 \quad \dfrac{\dfrac{}{\Gamma_1;\Delta_1 \vdash d_3 : D^{\Lambda}} \text{ T-A2}}{\Gamma_1;\Delta_1 \vdash \textbf{return } d_3; \rhd \Gamma_1;\Delta_1 \,|\, Cn_2} \text{ T-R}$$

where $\Delta_1 = \Delta, f_1 : \textbf{Fut}\langle D^{(\emptyset,\{\{u_4\},\{\text{HC,PT}\}\})}\rangle$ and $\Gamma_1 = d_3 : D^{(\emptyset,\{\{u_4\},\{\text{HC,PT}\}\})}$ and $\Lambda = (\emptyset, \langle\{u_4\}, \{\text{HC, PT}\}\rangle)$
$Cn_1 = \{A_{Cnstr}(\textbf{use}, \Lambda, \Delta_1), A_{Cnstr}(\textbf{collect}, \Lambda, \Delta_1)\}$ and $Cn_2 = Cn_1 \cup \{A_{Cnstr}(\textbf{transfer}, \Lambda, \Delta_1)\}$

■ **Figure 17** Type derivations for the statements of the method requestPatientInfo.

annotation (meaning that the set is empty). However, they are present in the type system annotation, $\textbf{Fut}\langle D^{(\emptyset,\{\{u_4\},\{\text{HC,PT}\}\})}\rangle$. No constraint is generated, since there is no personal information passed to the method.

- The second statement reads the future and assigns it to $d_3$ (see $\mathcal{D}_2$). Here we generate constraints due to the assignment to a local variable with personal data, which modify the type of the variable $d_3$ with the tag of the expected data from the future variable. Note that, the compatibility test for personal data is successful, since the type of $d_3$ in $\Gamma$ says that it is personal data.
- The final **return** statement (see $\mathcal{D}_3$) generates constraints since the returned expression has personal data. Looking at the definition in Figure 12, we can notice that we do not have to add the constraint $A_{Cnstr}(\textbf{use}, \Lambda, \Delta_1)$ since it was already in the set due to the previous statement.

From the derivations of Figure 17 we get the typing of the whole body as follow:

$$\dfrac{\mathcal{D}_1 \quad \dfrac{\mathcal{D}_2 \quad \mathcal{D}_3}{\Gamma;\Delta_1 \vdash d_3 := f_1.\textbf{get}; \textbf{return } d_3; \rhd \Gamma_1;\Delta_1 \,|\, Cn_1} \text{ T-SQ}}{\Gamma;\Delta \vdash \textbf{Fut}\langle D^{\#}\rangle \, f_1 := \textbf{this}.db!rpd(u_4); d_3 := f_1.\textbf{get}; \textbf{return } d_3; \rhd \Gamma_1;\Delta_1 \,|\, Cn_2} \text{ T-SQ}$$

Finally Rule (T-METH) is applicable and

$D^{\#} : (\_, \langle\{u_4\}, \{\text{Healthcare, Psychotherapy}\}\rangle) \text{ requestPatientInfo}(U\ u_4)\{ \ ..... \}, \langle Cn_2, \emptyset\rangle \ \textbf{OK in} \text{ HIS}.$

**Personal Data Enforcement** A type $T^{\#}$ in a variable indicates that it refers to personal data, whereas type $T$ indicates that it can be used freely (it will not require consent checking). In the initial type environments defined in Figure 16, the (program level) types ($T^{\#}$ and $T$) are transformed into the extended types needed in $\Gamma$. The # annotation (indicating that the data is personal) is transformed in the least restrictive tag annotation, whereas absence of annotation is transformed into $\epsilon$. Our type system enforces the restriction that a variable with a type which was initially tagged with $\epsilon$, is never assigned with personal data. So no data, which is meant to be personal can be assigned to a variable which is meant to be used freely. This is expressed by the following proposition.

**Proposition 1.** If $\Gamma;\Delta \vdash s \rhd \Gamma';\Delta' \,|\, \textbf{\textit{Cn}}$, then $\Gamma$ and $\Gamma'$ are such that $\Gamma(x) = T^{\epsilon}$ then $\Gamma'(x) = T^{\epsilon}$ and if $\Gamma'(x) = T^{\epsilon}$ and $x \in dom(\Gamma)$ then $\Gamma(x) = T^{\epsilon}$ .

```
entities {HospEnt, GPEnt, LabEnt}; purposes {Healthcare};
class Doctor(Lab lb) {
Unit collectSample(U u₁) {
    Int# patientRecord; // personal data
    Int patientReport; // non-personal data (public).
     // assigning personal data to a tagged variable
  patientRecord = 1234 tag ⟨{u₃}, {Healthcare}⟩
     // assigning Non-Personal data to personal variable
  patientRecord = patientReport ;

     // asynchronous call with incompatible parameter
  Fut<Unit> f₁ = lb!getSampleResults(u₁, patientRecord);
     // error: due to type incompatibility in the parameters of the call }}

class Lab( ) { ... Unit getSampleResults (U u₁, Int d) { /* calculate the sample results */ ...}
```

■ **Figure 18** Type Incompatibility Detection

*Proof.* By induction on type derivations. The side condition $\Lambda' = \epsilon \lor \Lambda \neq \epsilon$ of the rules that override the annotation of data types (where $\Lambda$ is the annotation in $\Gamma$) prevents deriving the annotation $\epsilon$ to a type not annotated by $\epsilon$. □

**Example 3.** Figure 18 demonstrates how type compatibility is analysed, ensuring that personal data is properly handled during program execution.

In the method collectSample, the local variable patientRecord is declared as personal data and patientReport is declared as non-personal data. Both variables are tracked in the local environment $\Gamma_0$. Initially

$$\Gamma_0 = \text{patientRecord} : \text{Int}^{⟨\emptyset,⟨\emptyset,\mathscr{P}⟩⟩}, \text{patientReport} : \text{Int}^\epsilon$$

giving to patientRecord the most permissive tag annotation.

- The first statement assigns personal data to patientRecord using the expression 1234 tag ⟨{u₃}, {Healthcare}⟩. According to Rule (T-ASSIGN-LOCAL), the assigned variable's type is updated to reflect the new tag annotation.
  The environment changes, and becomes

  $$\Gamma_1 = \text{patientRecord} : \text{Int}^{⟨u₃, \text{Healthcare}⟩}, \text{patientReport} : \text{Int}^\epsilon.$$

- In the second statement, assigning patientReport (non-personal) to patientRecord (personal) does not make the variable non-personal. Instead, patientRecord is updated with the most permissive tags, resulting in

  $$\Gamma_2 = \text{patientRecord} : \text{Int}^{⟨\emptyset,⟨\emptyset,\mathscr{P}⟩⟩}, \text{patientReport} : \text{Int}^\epsilon.$$

- In the next statement, asynchronous call to the method getSampleResults fails because patientRecord, passed as an actual parameter, does not satisfy the formal parameters' type requirements. The method expects a parameter of type $\text{Int}^\epsilon$, but patientRecord is personal with private tags. This type mismatch is caught by the *CmpTy* predicate, ensuring parameters passed to methods adhere to the type.

This example emphasizes the importance of *CmpTy* in enforcing type compatibility. It prevents privacy violations by ensuring personal data is used, passed, and assigned only when compatible with declared types, protecting sensitive information throughout the program's execution. The failures showcased illustrate the systematic and robust nature of TyPAOL in preserving personal data handling.

## 4 Reducing Data Privacy Compliance Checks at Runtime

The typed operational semantics of TyPAOL,[1] see Section 4.1, differs from that of P-AOL by checking user consent constraints at process activation, rather than at each access to tagged data. Information on the consents granted by the users is recorded in the *consent environment* $\Sigma$. This environment records, for each user, the set of actions that are permitted only when performed by a specific entity and for a specific purpose, i.e.,

$$\Sigma ::= \emptyset \mid \Sigma, u \mapsto \chi \quad \text{where} \quad \chi ::= \emptyset \mid \chi, a \mapsto \widehat{\langle \eta, p \rangle}$$

where in $\Sigma$ and $\chi$ there are no repetitions of users and actions, respectively.

Each active object is executing a process at a time and has a pool of processes waiting to be executed. At any time there may be several objects running. Messages generated by asynchronous method calls produce processes which are added to the process pool of the receiver object. In the operational semantics of P-AOL, when an object is idle, the activation simply makes the chosen process the running one. Instead in the typed operational semantics of Section 4.1 consent constraints are instantiated and checked at the time of method activation. To ensure that this is sound, since consent is a global shared resource, it is necessary to prevent interference from other concurrently running processes that may modify the consent of users checked by the current process, and dually that the current process does not modify the consent of users that are checked by some running process. So we add to any object the information on the users whose consent is modified and checked by its running process, and perform an interference check before activating the process. If no process in the process queue can be activated due to lack of the corresponding users' consent, no activity can happen in the object and it remains *idle*.

### 4.1 Typed Operational Semantics

The operational semantics of TyPAOL, similar to that of P-AOL (which was informally described in Section 2 and fully detailed in Appendix A), is defined by the reduction on configurations, $cn \ \Sigma \rightarrow cn' \ \Sigma'$, where $cn$ contains objects, messages and futures. An *object* is denoted by $o(\varphi, \gamma, q, \eta)^C$ where $\varphi$ are its fields, $q$ the pool of processes, $\eta$ is the entity associated to the object and

$$\gamma = idle \quad \text{or} \quad \gamma = \{\sigma \mid sts\}_\eta^f, \langle Um, Uc \rangle.$$

In the first case the *object is idle* and in the second the *object is running*, executing the statements *sts* where $\sigma$ is the local environment, $\eta$ is the entity accountable for the execution of the current code and $f$ is the future in which the result of the method will be deposited. Moreover, $\langle Um, Uc \rangle$ records in the execution of the current method the users whose consent is modified and the users whose consent is checked, respectively. In a message

$$m(o, o', \overline{val}, f, \eta'),$$

---

$$\text{(Bind-Mtd)}$$

$$\frac{I = \{i \mid val_i =_{t_i} v_i\} \quad \bigwedge_{i \in I} comply_T(\Sigma, \eta', t_i)}{B \; m(\overline{A}\,\overline{x}){:}\Lambda\{\,\overline{A'}\,\overline{y}; \; sts\}, \langle \boldsymbol{Cn}, \boldsymbol{Um} \rangle \; \textbf{OK in } C}$$

$$o(\varphi, \gamma, q, \eta)^C \; m(o, o', \overline{val}, f, \eta') \; cn \; \Sigma$$
$$\rightarrow o(\varphi, \gamma, q \; \langle\langle \overline{x} \mapsto \overline{val} \; \overline{y} \mapsto \overline{def(A')} \mid sts \rangle^f_{\eta'}, \langle \boldsymbol{Cn}, \boldsymbol{Um} \rangle\rangle_m, \eta)^C \; cn \; \Sigma$$

$$\text{(Activate)}$$

$$\frac{scheduler(q) = p = \langle\{\sigma|sts\}^f_{\eta'}, \langle \boldsymbol{Cn}, \boldsymbol{Um} \rangle\rangle \quad \forall \pi \in InstCnstr(\boldsymbol{Cn}, \langle \varphi, \sigma \rangle, \eta', \eta, \Sigma) \quad \pi \text{ holds}}{Um = \{\sigma(x)|x \in \boldsymbol{Um}\} \; Uc = Ck(\Lambda, \langle \varphi, \sigma \rangle)} $$
$$\frac{\forall o'(..., \langle Um', Uc' \rangle, ....)^{C'} \in cn \quad \neg conflict(\langle Um, Uc \rangle, \langle Um', Uc' \rangle)}{o(\varphi, idle, q, \eta)^C \; cn \; \Sigma \rightarrow o(\varphi, \{\sigma|sts\}^f_{\eta'}, \langle Um, Uc \rangle, (q \setminus p), \eta)^C \; cn \; \Sigma}$$

■ **Figure 19** Typed operational semantics of *method binding* and *process activation*.

$m$ is the method name, $o$ is the caller, $o'$ is the callee, $f$ is the future to which call's result is returned, $\eta'$ is the callee entity and $\overline{val}$ are the calls actual parameter values (can be seen in Figure 19). Finally in a configuration there can be futures, $f$ or $f(_t v)$. In the first case the instance of the method to which $f$ is bound to, is yet to be completed and in the second it contains its result. The reduction rules are the ones of P-AOL in which Rule (Activate) and Rule (Bind-Mtd) are replaced with the rules in Figure 19 and all the compliance check are removed, since we checks them only in Rule (Activate).

The updated Rules (Bind-Mtd) and (Activate) are given in Figure 19. Rule (Bind-Mtd) adds the process generated from the method call, including the constraints, to the process pool and checks compliance (as done for the same rule in P-AOL and reported in Appendix A). The local variables $\overline{y}$ are associated to default values. In case the variable is an object or a user its default value is the (unique) fresh identifiers that will be returned when the object/user will be created. Observe that we have modified the runtime syntax of processes in the waiting process pool, and now they also include the gathered constraints. In Rule (Activate), we check the constraints gathered by the type system for the whole method body of the process being activated and also that there will be no interference on the users's consents accessed and/or modified during the execution of the process. We gather the checked users using *Ck* and check interference using the predicate *conflict*, further details are given in Definition 3. In this rule, $InstCnstr(\boldsymbol{Cn}, \langle \varphi, \sigma \rangle, \eta', \eta, \Sigma_0)$ *instantiates the constraints in* $\boldsymbol{Cn}$ producing $act \in \mathscr{A}(\Sigma, \eta', \langle \widehat{u}, \widehat{p} \rangle)$ from $A_{Cnstr}(act, \Lambda, \Delta)$ given that $act \neq$ **store**, and it produces **store** $\in \mathscr{A}(\Sigma, \eta, \langle \widehat{u}, \widehat{p} \rangle)$ from $A_{Cnstr}(\textbf{store}, \Lambda, \Delta)$. This implies that all the actions except for **store** must be authorised for the accountable entity, whereas **store** should be authorised for the entity associated with the callee. To do so, we need the information on the runtime environment.

The *values of the fields and local variables*, $\langle \varphi, \sigma \rangle$, are used to get the $\langle \widehat{u}, \widehat{p} \rangle$ from the tag annotation $\Lambda$: $\langle \widehat{u}, \widehat{p} \rangle$ is obtained from $\Lambda$, by replacing (i) *vr* with the tags of the values associated with the parameter or the field named *vr*, and (ii) user variables with user identifiers. Then, upon substitution we compose the resulting tags with the

$$plcy(R, \chi) = \begin{cases} \chi & \text{if } R = \_ \\ \chi'[\chi'(a) \mapsto \chi'(a) \cup \{(\eta, p)\}]_{a \in \widehat{a}} & \text{if } R = R' + \rho \wedge \chi' = plcy(R', \chi) \\ \chi'[\chi'(a) \mapsto \chi'(a) \setminus \{(\eta, p)\}]_{a \in \widehat{a}} & \text{if } R = R' - \rho \wedge \chi' = plcy(R', \chi) \end{cases}$$

■ **Figure 20** Changes to policy $\chi$ due to $R$

$\oplus$ operator. Here the *tg* operator gives the tags of a variable. The *instance of a tag annotation w.r.t. to fields and local variables* $\langle \varphi, \sigma \rangle$, $InstAnn(\Lambda, \langle \varphi, \sigma \rangle)$, is defined by:

$$InstAnn(\epsilon, \langle \varphi, \sigma \rangle) = \epsilon$$
$$InstAnn((\widehat{vr}, \widehat{z}, \widehat{p}), \langle \varphi, \sigma \rangle) = \oplus_{\text{this}.x \in \widehat{vr}} tg(\varphi(x)) \oplus_{y \in \widehat{vr}} tg(\sigma(y)) \oplus \langle \{\sigma(z) \mid z \in \widehat{z}\}, \widehat{p} \rangle$$

The typing rules enforce uniqueness of association between variables and users in the body of method. The consent environment $\Sigma$ is obtained from the *initial consent environment*, $\Sigma_0$, and $\Delta$ containing the modifications of the consent per each user in $dom(\Delta)$. To capture this, Figure 20 defines how a *policy expression $R$ modifies a policy map $\chi$*.

The instantiation of constraints is defined by the following.

**Definition 2.** *The instantiation of the constraints in $\boldsymbol{Cn}$, $InstCnstr(\boldsymbol{Cn}, \langle \varphi, \sigma \rangle, \eta_0, \eta_1, \Sigma)$, is defined by cases on the shape of the constraints in $\boldsymbol{Cn}$:*

- $A_{Cnstr}(act, \Lambda, \Delta) \in \boldsymbol{Cn}$ *produces* $act \in \mathscr{A}(\Sigma[\Delta], \eta_0, \langle \widehat{u}, \widehat{p} \rangle)$ *if* $act \neq \boldsymbol{store}$
- $A_{Cnstr}(\boldsymbol{store}, \Lambda, \Delta) \in \boldsymbol{Cn}$ *produces* $\boldsymbol{store} \in \mathscr{A}(\Sigma[\Delta], \eta_1, \langle \widehat{u}, \widehat{p} \rangle)$

*where*

- $\langle \widehat{u}, \widehat{p} \rangle = InstAnn(\Lambda, \langle \varphi, \sigma \rangle)$ *and*
- $\Sigma[\Delta] = \Sigma \begin{cases} [u \mapsto plcy(R, \chi)]_{\Sigma(u) = \chi \wedge \Delta'(u) = \mathsf{U}\langle R \rangle} \\ [u \mapsto plcy(R, \emptyset)]_{u \notin dom(\Sigma) \wedge \Delta'(u) = \mathsf{U}\langle R \rangle} \end{cases}$

  *and* $\Delta' = [u : \mathsf{U}\langle R \rangle \mid \Delta(x) = \mathsf{U}\langle R \rangle \wedge \sigma(x) = u]$ *and plcy is defined in Figure 20.*
  $\Delta'$ *takes into account the references to users and substitutes the variables referring to users with the user they refer to.*

We now define the users whose consent is checked by $\boldsymbol{Cn}$ and the interference predicate between between users.

**Definition 3.** - $Ck(\Lambda, \langle \varphi, \sigma \rangle) = \widehat{u}$ *where* $\langle \widehat{u}, \widehat{p} \rangle = InstAnn(\Lambda, \langle \varphi, \sigma \rangle)$
- *The predicate* $conflict(\langle Um, Uc \rangle, \langle Um', Uc' \rangle)$ *is defined by*
$$(Um \cap Um') \cup (Um \cap Uc') \cup (Um' \cap Uc) \neq \emptyset$$

A process is activated only if its execution does not interfere with any process currently running in the system. Interference arises in two cases, 1) if another running process is modifying or checking the consent of any user that the current process intends to modify, or 2) if another process is modifying the consent of any user that the current process intends to check. This ensures that the consent assumptions verified at activation remain valid throughout execution.

**Example 4.** Building upon Example 2, where we type-check a method in our hospital information system, we now extend the analysis to include method binding and process activation to showcase the typed operational semantics. Using the derivations provided in Figure 17, the entire method body is typed by applying Rule (T-Meth). At runtime, a method call to requestPatientInfo adds a process to the pool, including all associated constraints. Specifically, at Rule (Activate), these constraints, generated by the type system, are checked against the runtime environment.

The constraints generated in Example 2 are as follows:

$$Cn_2 = \{A_{Cnstr}(\textbf{use}, \Lambda, \Delta_1), A_{Cnstr}(\textbf{collect}, \Lambda, \Delta_1), A_{Cnstr}(\textbf{transfer}, \Lambda, \Delta_1)\}$$

where $\Lambda = (\emptyset, \langle\{u_4\}, \{HC, PT\}\rangle)$. At runtime, using the environment $\langle\varphi, \sigma\rangle$, the instantiation of the tag annotation with respect to fields and local variables is computed as:

$$InstAnn((\emptyset, \langle\{u_4\}, \{HC, PT\}\rangle), \langle\varphi, \sigma\rangle) = (\emptyset, \langle\{bob\}, \{HC, PT\}\rangle)$$

From Definition 2, the instantiation of constraints is $InstCnstr(\boldsymbol{Cn}, \langle\varphi, \sigma\rangle, \eta_0, \eta_1, \Sigma)$ and for the method requestPatientInfo we have:

$Cn_2 = \{A_{Cnstr}(\textbf{use}, \Lambda, \Delta_1), A_{Cnstr}(\textbf{collect}, \Lambda, \Delta_1), A_{Cnstr}(\textbf{transfer}, \Lambda, \Delta_1)\}$, $\eta_0 = \mathsf{GpEnt}$, $\eta_1 = \mathsf{HospEnt}$ and since $\Sigma$ does not change in the method body, here $\Sigma = \Sigma_0$ which is initial consent from main method.

Given the shape of the constraints, *instantiateConstr* produces the following,

- $A_{Cnstr}(\textbf{use}, \Lambda, \Delta_1)$ produces $\textbf{use} \in \mathscr{A}(\Sigma_0, \mathsf{GpEnt}, \langle\{bob\}, \{HC, PT\}\rangle)$
- $A_{Cnstr}(\textbf{collect}, \Lambda, \Delta_1)$ produces $\textbf{collect} \in \mathscr{A}(\Sigma_0, \mathsf{GpEnt}, \langle\{bob\}, \{HC, PT\}\rangle)$
- $A_{Cnstr}(\textbf{transfer}, \Lambda, \Delta_1)$ produces $\textbf{transfer} \in \mathscr{A}(\Sigma_0, \mathsf{GpEnt}, \langle\{bob\}, \{HC, PT\}\rangle)$

Additionally, using Definition 3, we compute the set of users whose consent is accessed by the method: $Ck(\Lambda, \langle\varphi, \sigma\rangle) = \{bob\}$ (*Uc*), and from Example 2, we have $Um = \emptyset$. To safely activate the process, the runtime system using the *conflict* predicate, checks that no other object is currently executing a process that modifies or checks Bob's consent in a way that would conflict with this method. If all instantiated constraints hold and no interference is detected, the process is safely activated and begins execution. This demonstrates how the constraints generated by the type system during compilation are dynamically checked against the runtime information at the method activation point, ensuring compliance with user-defined consent.

### 4.2 Soundness of the Type System for the Typed Operational Semantics

In this section we outline the soundness result for our type system. This is expressed by stating that: if a program *PR* is well-typed, any non-idle object in a configuration is either waiting for a future to be fulfilled or can make a reduction step.
This implies that,

- on one side, a running object *does not get stuck* for errors such as: expressions containing undefined operators, or the test of a conditional statement not resulting in a boolean value or the classical object oriented error which is calling a method which is not defined for the class of the object it is called on, and
- on the other, *all the checks related to the handling of personal data* can be avoided, since we can prove that they *will always succeed*.

The latter is obtained at the expense of anticipating these checks before the activation of the process of the running objects.

Let $\rightarrow^*$ denote any number of reduction steps $\rightarrow$ and *CN* be the global configuration consisting of a configuration *cn* and a global consent $\Sigma$, i.e., $CN = cn\ \Sigma$. In the following definition with $\perp$ we denote a default entity.

For a configuration, *reachability* from the beginning of a *well-typed PR*, is defined as follows:

**Definition 4** (Reachable Configuration)**.** *Let PR = E P $\overline{CL}$ $\{\overline{A}\,\overline{x}$; sts$\}$ be well-typed. The configuration CN is* reachable *from PR, if*

$$main(\emptyset, \{\overline{x} \mapsto \overline{def(A)}, \text{\textcolor{purple}{this}} \mapsto main \mid sts\}_\perp, \langle \emptyset, \emptyset \rangle, \perp)^{Main}\ \emptyset\ \to^*\ CN$$

Objects running some process in a reachable configuration can make a reduction step unless they are waiting for a future, i.e., the compliance predicates on personal data are always satisfied.

**Theorem 1** (Progress for running objects)**.** *Let $o(\varphi, \{\sigma \mid sts\}_{\eta'}^f, \langle Um, Uc \rangle, q, \eta)^C\ CN$ be reachable from a well-typed PR, then for some CN$'$*

$$o(\varphi, \{\sigma \mid sts\}_{\eta'}^f, \langle Um, Uc \rangle, q, \eta)^C\ CN \to CN'$$

*or sts has the form $y := x.\text{\textcolor{purple}{\textbf{get}}}$ and $\sigma(x) = f'$ and $f' \in CN$.*

The proof of the theorem, given at the end of Appendix C, relies on the definition of *well-typed configuration* (Figure 30), and some *subject reduction* lemmas. In particular, subject reduction is proved for the reduction of a process in a running object (Lemmas 3, 4 and 5) and also for the rules for binding and activating methods (Lemma 6). *Well-typed configurations* contain running object whose statements are well-typed w.r.t. some typing environment $\Gamma$ and $\Delta$. The typing judgment for statements is of the shape $\Gamma; \Delta \vdash sts \triangleright \Gamma'; \Delta' \mid \ldots$ where $\Gamma'$ and $\Delta'$ track the changes to the tag annotations of personal variable types in $\Gamma$ and to reference environment in $\Delta$ due to the execution of *sts*. In addition, in a well-typed configuration there should not be running object with conflicting modified and checked user's consents.

*Subject reduction*, for running objects, asserts that reduction of an object which is well-typed w.r.t. $\Gamma$ and $\Delta$ produces a configuration which is well-typed w.r.t. $\Gamma'$ and $\Delta'$. That is, on one side tagged values are modified in accord to the tag annotation of the variables that refer to them and, on the other, that the changes to the consent of users which are needed for the execution of the method are tracked by the changes to the policy expressions of the associated variables. Since in the typed semantics we anticipates the checking of constraints at the beginning of the method that should run, we have to prove that during the execution of the statements *sts* there cannot be a running object which modifies consents of users that are either modified or checked during the execution of *sts*. This is achieved by avoiding

1. intra-method aliasing to users, and

2. inter-method interference between running objects.

The first property is enforced by the type system. In particular, Rule (T-NEW-USER) does not allow aliasing between two local variables and Rule (T-ASYN-CALL) does it for formal parameters. The second property is enforced by the interference check in Rule (ACTIVATE). With these properties we can prove the crucial lemma for the proof of progress, which asserts that reachable configurations are well-typed (Lemma 9).

As stated at the beginning of the section, Theorem 1 allows us to remove the compliance checks from the rules of the typed operational semantics.

## 5    System Integration Considerations

While our formalism provides guarantees for data privacy compliance in a language-based model, implementing such a system in practice needs careful strategies. This section outlines how the proposed semantics and constraints could be embedded in real-world systems and discusses possible implementation mechanisms.

**Consent Access as Read/Write Locks**    The operational semantics ensures that no process interferes with the consent state assumed by another process and in a real language, this could be implemented, instead of checking all running objects, by using read/write locks on users. These locks encode the conflict discipline defined by our interference predicate and for a process to be activated, it would need to acquire the locks required for its execution and release them at the end of it. This aligns with runtime isolation guarantees from type-based memory models to avoid data races. In a real system, consent consistency could be enforced by an interference-aware scheduler using read/write locks on users. Which is similar to Rust's ownership model, the focus here is runtime enforcement of consent consistency, instead of memory safety.

**Consent Validation and User Authentication**    The type system statically determines, for each method, the set of users whose consent is intended to be modified. It tracks the exact changes in consent (i.e., add or remove) in our type environment upfront. Although not part of our core formalism, the typed operational semantics can be extended with an explicit user authenticator to ensure that any changes to the user's consent should be explicitly authorised by that user to avoid invalid consent being added to the system. This could be implemented, in a real system, by integrating an authentication hook to users, involving consent dialogues which at the time of activation check if each consent modification tracked by the type system is user approved. The type system can act as a static extractor of consent dependencies, which can be used by the consent authenticator.

**Type System Integration**    TyPAOL's static-dynamic model can be integrated in modern languages with effect systems or type-based capability tracking. Systems like Scala, Haskell, or TypeScript could support consent and tag annotations via type refinements or custom type-checking passes [34, 48]. Runtime enforcement could be layered on top using actor-based or task-based execution frameworks (e.g., Akka [33], Orleans [6]).

## 6    Related Work

This paper explores a relatively recent area of research that combines the intersection of GDPR, privacy, and programming language theory. It examines the static and dynamic aspects of personal data flow from a programming language perspective and mainly focuses on privacy philosophies like privacy by design, policy enforcement, policy specification, and privacy-enhancing technology. Below, we relate our work against various relevant approaches.

**Information Flow Control Approaches**    TyPAOL differs from traditional IFC methods static, dynamic, hybrid, and decentralised by integrating user consent enforcement and GDPR compliance into programming language semantics. Unlike static IFC, which enforces policies at compile time and may reject safe programs [12, 26, 35, 50], TyPAOL combines static type inference with runtime constraint checking, ensuring privacy while allowing dynamic consent changes. It prevents middle of the execution of a method failures, a limitation of Dynamic IFC [3, 4, 14, 36, 49], and optimises hybrid IFC by limiting privacy checks to method binding and activation [23, 24, 38], reducing execution overhead. Unlike security-centric decentralised IFC models such as DStar, HiStar, LIO, and Flume [21, 42, 52, 53], TyPAOL enforces user consent at both static and runtime levels, ensuring GDPR compliance while preventing unauthorised data usage in domains like healthcare, social media, and finance. Recent advances in hybrid IFC such as, type-driven gradual security [45] enhances flexibility by combining static and dynamic checks, mainly focusing on security levels. Similarly, hybrid IFC for low-level code [15] strengthens value-dependent enforcement but operates at a system level. In contrast, TyPAOL advances the state of the art by embedding user consent and GDPR compliance directly into the type system and runtime semantics, enabling dynamically adjustable privacy control at the application level, which is missing in prior work that focused solely on confidentiality and integrity.

**Enforcement of Privacy Policies**   Language-based enforcement of privacy policies has been investigated before. In [18] the authors model and verify privacy policies mainly specified in P3P, for the JIF programming language. JIF provides static information flow checking through a type system-based decentralised label model, allowing programmers to annotate variables, methods, and classes with purpose and entities. However, this approach does not assure the validity of personal data handling and overlooks the dynamic aspects of users' consent. In [1], the authors devise a type system for the messages exchanged between black box components. They guarantee that privacy-related properties between components must hold if all messages exchanged are well-typed. Privacy compliance checking for active object languages has been addressed from a static perspective in [44], where access rights to personal data are interpreted as an extended notion of read/write access, and from a dynamic perspective [5, 20, 43], resulting in runtime checks that include purely static constraints. In this paper, we combine the benefits of static and dynamic approaches.

**Effects and Coeffect Systems**   Coeffect systems are a form of type systems modelling how execution needs to use external resources that are bound to variables [28]. They are, in a sense, the dual of effect systems in which types are enriched to carry information on the side effects of the execution, [27]. In our type system, on one side, we impose restrictions on the use of variables expecting to hold personal data and prevent aliasing of references and, on the other side, we trace the effect of the changes to the consent environment and also changes to the tag annotations of variables expecting to hold personal data. So we can see the first as sort of coeffects and the second as effects. We could consider our tag annotations of types as grades; however, we do not elaborate an algebra for tag annotations. We defined some operators and a partial order and used their properties. Taking a more abstract approach to tracing tag annotations is part of our future work. We are not aware of results in this direction.

**Gradual Typing**    Among techniques for integrating static and dynamic typing within a single language, we have gradual typing [39], where behavioural types are introduced into existing systems gradually so that the typed and less typed programs can interoperate using suitable type casts at the interfaces. Gradual types emerge from dynamic typing, whereas our work does not address dynamic typing, but it focuses on inferring program points for runtime instrumentation (e.g, at method activation) based on the static typing results (e.g., collection of constraints).

**Runtime Monitoring of Properties**    Static analysis and runtime monitoring for security and privacy-related behaviours of a system has been explored previously. In [10], the authors propose an approach that combines static analysis techniques and runtime verification to efficiently monitor data flow related hyper-properties, including security for information flow. By statically analysing the code, they identify potential violations of such properties and then use runtime verification to confirm these violations during system execution. This approach differs from ours as we use annotation-based code analysis instead of runtime monitoring. Similarly, in [2] and [19], the authors propose runtime monitoring for GDPR specifications to automate compliance checks. Log-based monitoring checks for privacy violations and infers actions on data from the log, and it adds runtime overhead, which contrasts with our approach where we infer actions before execution. Various runtime verification techniques [17, 32, 51] depend on monitoring the state and behaviour of systems and then analysing the gathered data to determine whether design-level properties are met. However, as far as we know, there has not been an effort to integrate dynamic design-level properties, such as consent, into static analysis while also tracking changes. Our approach can help to avoid the overhead associated with consent monitoring at runtime.

## 7   Conclusion and Future Work

In this paper, we present an approach to enhance an active object language with built-in abilities to deal with data privacy compliance checks. The proposed approach considers a privacy-aware language with a careful integration of GDPR essentials, such as privacy policies, individual users' consent, data handling procedures, and lawful processing, which together constitute the basis for checking data privacy compliance. To guarantee data privacy compliance checks, we first propose a type system and type inference for the language to track the flow of personal data and generate constraints for privacy checks. Achieving data privacy compliance checks solely through static techniques is impossible due to the dynamic nature of consent and data flow. Therefore, we employ the constraints generated from the static method within a runtime context to ensure data privacy compliance.

P-AOL explores the integration of GDPR principles into a distributed object-oriented language, by extending the language with specific constructs that make the language privacy-aware. It could be interesting to explore whether similar extensions could be done to mainstream languages via libraries or middleware extensions, providing a foundation for further data privacy analysis. An important aspect to consider here is that the active object setting provides concurrency together with explicit encapsulation and data-race freedom, allowing us to focus on data privacy checks. For instance, at runtime, processes that violate privacy constraints are suspended and postponed, as privacy consent can change dynamically. Addressing such concurrency issues in different concurrency models is beyond the scope of this paper. Additionally, a privacy analysis tool using TyPAOL 's type system and consent management constructs could be developed to perform static detection of non-compliant data flows and runtime validation of GDPR constraints, providing a practical compliance framework for real-world systems.

The semantics of the language addresses some fine-grained personal data handling concepts for GDPR in the transition rules. However, crucial concepts like data deletion (pertaining to the right to be forgotten) and data retention time (related to storage limitation) are omitted, due to their temporal implications in the language and its analysis. Extending the semantics and static analysis to include these concepts would be interesting, enhancing the scope of the compliance checks.

The semantics of the language formalises the transfer of rights between DCs and DPs by the notion of *accountable entity*. Extending compliance checks to include restrictions on processing rights for processors and sub-processors by DCs would be interesting. GDPR highlights contractual agreements between DCs and DPs, along with the processor's responsibility to the DC. Hence, we may need to add privacy policies for DCs to restrict the information exchange between controllers and (sub)processors, e.g., by keeping track of the approved processors for each controller.

While TyPAOL provides strong guarantees under controlled assumptions, we acknowledge that practical deployment needs careful strategies. Specifically, our current system assumes limited aliasing and user authenticator. Ensuring these properties in distributed systems will require further research. In particular, maintaining object uniqueness and enforcing aliasing constraints in practice may need mechanisms such as centralised user identity management, consent interface, distributed consistency protocols, or runtime uniqueness checks [7, 46]. Although these concerns lie outside the current scope, but they are important directions for future work.

Type systems targeting privacy properties can highly benefit from behavioural types, which also specify how programs compute in a distributed setting (i.e., the structure of computations) thus giving more description of the global behaviour, whose semantics is based on the interactive behaviour of various parties. Since our setting exhibits interaction patterns between actors, it can be captured using e.g., session types, where type abstractions can be done on behaviours to ensure privacy-related properties.

Finally, we know that many legal frameworks share core principles of GDPR such as purpose limitation, consent-based data handling, and user-centric control. By parameterising the consent actions and purpose tags, TyPAOL could be adapted to jurisdictions where the terminology or scope of data protection differs.

**Acknowledgements**   This work was partially funded by the MUR project "T-LADIES" (PRIN 2020TL3X8X) and has the financial support of the University of Eastern Piedmont. The authors are strongly indebted to the anonymous referees for their constructive remarks, which helped us improve the submitted version of the paper.

## A  Operational Semantics of P-AOL

In this section, we give the formal operational semantics of P-AOL. It is a slight variant of the one presented in [5]. We first introduce the runtime syntax and some functions and operations on tags and consent. We then present the rules for the evaluation of expressions, and the ones for the evaluation of statements. The semantics of expressions is given in a big-step style whereas the one of statements is a small-step semantics.

$$
\begin{array}{llll}
CN & ::= & cn\,\Sigma & \\
cn & ::= & obj\,cn \mid msg\,cn \mid fut\,cn \mid \epsilon & \\
obj & ::= & o(\varphi, \gamma, q, \eta)^C & \\
\sigma, \varphi & ::= & \emptyset \mid \sigma[x \mapsto val] & \\
\gamma & ::= & \{\sigma | sts\}^f_\eta \mid idle & \\
q & ::= & \overline{\{\sigma | sts\}^f_\eta} &
\end{array}
\qquad
\begin{array}{llll}
msg & ::= & m(o, o', \overline{val}, f, \eta) \\
fut & ::= & f \mid f(_t v) \\
\Sigma & ::= & \emptyset \mid \Sigma[u \mapsto \chi] \\
\chi & ::= & \emptyset \mid \chi[a \mapsto \widehat{(\eta, p)}] \\
val & ::= & u \mid o \mid f \mid _t v \mid v \\
t & ::= & \epsilon \mid \langle \widehat{u}, \widehat{p} \rangle \\
sts & ::= & \mathbf{unit} \mid s; sts \mid \mathbf{return}\,\alpha
\end{array}
$$

■ **Figure 21**  Runtime Syntax of TyPAOL.

**Runtime Syntax**   Figure 21 shows the runtime syntax of TyPAOL. A global *configuration CN* consists of a configurations $cn$ and a global user-specific consent $\Sigma$. Where $cn$ is a sequence of runtime elements: objects, invocation messages and futures. For simplicity, classes are not represented explicitly in the semantics, but may be seen as static look-up tables of object layout and method definitions.

An *object obj* is a term $o(\varphi, \gamma, q, \eta)^C$, where $o$ is the object's identifier, $\varphi$ is the state of the object and consists of the binding of the object's fields to values, $\gamma$ is the process currently being executed which may be idle, $q$ a *pool of processes* waiting to be scheduled for execution, $\eta$ is the entity associated with the object, and $C$ is the class of the object. We will sometimes omit $C$ when this is not relevant. A *process* $\{\sigma | sts\}^f_\eta$ consists of a local state $\sigma$ of local variable bindings, a list *sts* of statements, the entity $\eta$ that made the call, and a reference to the future $f$ that will hold its result, or it is *idle*. Both $\sigma$ and $\varphi$ are mappings between names of variables and fields to the (runtime) values. We use the overwriting operation $\_[\_\mapsto\_]$ for mappings, where e.g., $\sigma[x \mapsto val]$ denotes the overwriting of the value associated to $x$ in $\sigma$ with $val$, if $x$ is not in the domain of the map, then the full entry is added to it. An *invocation message* is a term $m(o, o', \overline{val}, f, \eta)$, consisting of a method name $m$, $o$ the object callee, $o'$ the object caller, $\overline{val}$ the call's actual parameter values, $f$ the future to which the call's result is returned and $\eta$ the entity associated with the object that called the method. A future *fut* is either an identifier $f$ or an identifier and a return value $_t v$. The consent $\Sigma$ is a mapping from users $u$ to policy maps $\chi$. The policy $\chi$ is a mapping that records for every action $a$, a set of pairs of entity and purpose, $a \mapsto \widehat{(\eta, p)}$, meaning that the entity $\eta$ is allowed to perform action $a$ for the user $u$ for purpose $p$. The empty policy $\emptyset$ states that no entities or purposes are allowed for any action. Values can be references to users and objects or tagged and untagged data values. We use the metavariable $v$ for single untagged values and untagged tuples of values. A runtime tag $t$ can either be empty or $\langle \widehat{u}, \widehat{p} \rangle$, a sets of users and a set of purposes. Values tagged by an empty tag $_\epsilon v$ are equivalent to untagged values $v$ and are always treated as non-personal data. To represent the results of the evaluation of a statement, we added the value **unit** as the (unique) value of type *Unit*, to indicate computation completion in a method without a return statement. To simplify the operational semantics and type rules, we replace $\epsilon$ with **unit** in sequences of statements *sts*, see Figure. 9.

**Tags and Consent**  Tags are combined using the $\oplus$ operator (see [5]) defined as follows:

$$t_1 \oplus t_2 = \begin{cases} t_i & \text{if } t_j = \epsilon \wedge t_i \neq \epsilon \quad i,j \in \{1,2\} \\ \langle \widehat{u} \cup \widehat{u}', \widehat{p} \cap \widehat{p}' \rangle & \text{if } t_1 = \langle \widehat{u}, \widehat{p} \rangle \wedge t_2 = \langle \widehat{u}', \widehat{p}' \rangle \\ \epsilon & \text{otherwise} \end{cases}$$

Intuitively, $t_1 \oplus t_2$ combines the privacy restrictions expressed by $t_1$ and $t_2$. For non-empty tags, since $\langle \widehat{u}, \widehat{p} \rangle$ tags a data which is personal for all users in $\widehat{u}$ and can be used for at least a purpose in $\widehat{p}$, we unify the users, so the result is personal for all of them, and intersects the purposes, so every user should explicitly give consent for the resulting purposes. Given a non empty tag, we extract its users and purposes by

$$\mathscr{U}(t) = \{u \mid t = \langle \widehat{u}, \widehat{p} \rangle \wedge u \in \widehat{u}\} \qquad \mathscr{P}(t) = \{p \mid t = \langle \widehat{u}, \widehat{p} \rangle \wedge p \in \widehat{p}\}$$

The preorder $\sqsubseteq$ on tags defined by

$$t_1 \sqsubseteq t_2 \quad \text{iff} \quad \mathscr{U}(t_2) \subseteq \mathscr{U}(t_1) \ \wedge \ \mathscr{P}(t_1) \subseteq \mathscr{P}(t_2)$$

expresses that values tagged by $t_1$ are more restrictive than values tagged by $t_2$, i.e., the less users the less requirements, and the more purposes, the better (making more processing possible). It is easy to see that $t_1 \oplus t_2 \sqsubseteq t_i, i \in \{1,2\}$ and that $t_1 \oplus t_2$ is the greatest lower bound of $t_1$ and $t_2$.

Similarly, there is a preoder over consent by

$$\Sigma_1 \sqsubseteq \Sigma_2 \quad \text{iff} \quad \forall u \ \Sigma_1(u) = \chi_1 \implies \Sigma_2(u) = \chi_2 \ \wedge \ \forall a \ \chi_1(a) \subseteq \chi_2(a)$$

That is, if $u$ allows action $a$ to be done by entity $\eta$ for the purpose $p$ according to $\Sigma_1$, the same has to hold according to $\Sigma_2$.

The function $\mathscr{A}(\Sigma, \eta, t)$ returns a set of actions, it extracts for any given tag $t$ and entity $\eta$, the *allowed actions* granted by the users in the consent $\Sigma$. If the tag $t$ is empty, all actions are granted and if $\mathscr{P}(t) = \emptyset$ no action is granted.

$$\mathscr{A}(\Sigma, \eta, t) = \begin{cases} \bigcap_{u \in \mathscr{U}(t)} \{a \mid \Sigma(u) = \chi \ \wedge \ \exists p \in \mathscr{P}(t) \quad (\eta, p) \in \chi(a)\}, & \text{if } t \neq \epsilon \\ \{\textbf{use}, \textbf{collect}, \textbf{transfer}, \textbf{store}\} & \text{otherwise} \end{cases} \quad (1)$$

The preorder on tags and consent relates in a natural way to their allowed actions. That is, *the more restrictive the tags or the consents are, the less actions are allowed*, which in turn means that the privacy conditions of the operational semantics are violated more often, i.e.,

$$\begin{aligned} t_1 \sqsubseteq t_2 & \quad \Rightarrow \quad \mathscr{A}(\Sigma, \eta, t_1) \subseteq \mathscr{A}(\Sigma, \eta, t_2) \\ \Sigma_1 \sqsubseteq \Sigma_2 & \quad \Rightarrow \quad \mathscr{A}(\Sigma_1, \eta, t) \subseteq \mathscr{A}(\Sigma_2, \eta, t) \end{aligned} \qquad .$$

**Evaluation of Expressions**  The evaluation of expressions is given by the judgement $e \mid \langle \varphi, \sigma \rangle \Downarrow val$ of Figure 22, where $\varphi$ and $\sigma$ associate values to the object fields and the local variables, respectively (see Figure 21), and they are the context for evaluating the expression $e$ to *val*. Expression evaluation terminates in a value unless a local variable, field, or operand type is undefined or incorrect. Most of the rules are obvious

<div align="center">

(VAL)   (VAR)   (FIELD)

$val \mid \langle \varphi, \sigma \rangle \Downarrow val$   $x \mid \langle \varphi, \sigma \rangle \Downarrow \sigma(x)$   $\textbf{this}.x \mid \langle \varphi, \sigma \rangle \Downarrow \varphi(x)$

(E-TAG)

$$\frac{e \mid \langle \varphi, \sigma \rangle \Downarrow_t v \quad t' = t \oplus \widehat{\langle \sigma(x), \hat{p} \rangle}}{e \; \textbf{tag} \; \langle \hat{x}, \hat{p} \rangle \mid \langle \varphi, \sigma \rangle \Downarrow_{t'} v}$$

(OP)

$$\frac{e_1 \mid \langle \varphi, \sigma \rangle \Downarrow_{t_1} v_1 \quad e_2 \mid \langle \varphi, \sigma \rangle \Downarrow_{t_2} v_2}{e_1 \; op \; e_2 \mid \langle \varphi, \sigma \rangle \Downarrow_{t_1 \oplus t_2} (v_1 \; \underline{op} \; v_2)}$$

(TUPLE)

$$\frac{e_i \mid \langle \varphi, \sigma \rangle \Downarrow v_i \quad i \in \{1, .., n\}}{(e_1, \ldots, e_n) \mid \langle \varphi, \sigma \rangle \Downarrow (v_1, \ldots, v_n)}$$

(SELECT)

$$\frac{e \mid \langle \varphi, \sigma \rangle \Downarrow (v_1, \ldots, v_n) \quad i \in \{1, .., n\}}{e.i \mid \langle \varphi, \sigma \rangle \Downarrow v_i}$$

</div>

■ **Figure 22** Evaluation of expressions and sequences of expressions.

$$comply_U(\Sigma, \eta_{acc}, t) \quad = \quad \textbf{use} \in \mathscr{A}(\Sigma, \eta_{acc}, t)$$
$$comply_C(\Sigma, \eta_{acc}, t) \quad = \quad comply_U(\Sigma, \eta_{acc}, t) \wedge \textbf{collect} \in \mathscr{A}(\Sigma, \eta_{acc}, t)$$
$$comply_T(\Sigma, \eta_{acc}, t) \quad = \quad comply_U(\Sigma, \eta_{acc}, t) \wedge \textbf{transfer} \in \mathscr{A}(\Sigma, \eta_{acc}, t)$$
$$comply_S(\Sigma, \eta_{acc}, \eta_{callee}, t) \quad = \quad comply_C(\Sigma, \eta_{acc}, t) \wedge \textbf{store} \in \mathscr{A}(\Sigma, \eta_{callee}, t)$$

■ **Figure 23** Comply predicates to check users' consent in P-AOL.

Here we comment only the relevant ones. In Rule (E-TAG), the tag of the result of the evaluation $t$, from the expression $e$ is combined with the one provided in the **tag**-construct $\langle \hat{x}, \hat{p} \rangle$, after substituting its variables $\hat{x}$ with the references to the users in the local environment $\sigma$. For binary operations $op$, Rule (E-OP), we let $\underline{op}$ capture the actual arithmetic and logic operators that can directly be applied over values. If the operands are evaluated to data values $_{t_1} v_1$ and $_{t_2} v_2$, the tags $t_1$ and $t_2$ are combined according to the operator $\oplus$ and the values $v_1$ and $v_2$ according to $\underline{op}$.

**Compliance of Consent** Personal data for which a user $u$ must have given consent to do some actions for the purposes in $\hat{p}$ are tagged by $\langle \hat{u}, \hat{p} \rangle$ with $u \in \hat{u}$. In order to handle such data in accord to consent of the user, we define the runtime predicates of Figure 23. The predicate $comply_U$ is checked when an expression involving the data is evaluated and says that the user must have granted *use* of the data to the accountable entity. If the result of the expression is then assigned to a local variable the user must have also allowed the right to *collect* the data (predicate $comply_C$), instead if it is passed as argument to a method the user must have granted the right to *transfer* the data (predicate $comply_T$). Finally in order to store the result expression in a field of the current object, which is acting as the entity $\eta_{callee}$, in addition to the right of using and collecting the data for the accountable entity, the user must have granted to the entity $\eta_{callee}$ the right to *store* it in a field.

If $t = \epsilon$, then $\mathscr{A}(\Sigma, \eta, t)$ is the set containing all the actions, so the consent predicates are always true.

**Rules of the Operational Semantics** The semantics is specified by a reduction relation on configurations of the form $CN \to CN'$. To simplify the definition of the reduction rules via $\rightarrow$, we use the reduction step $\rightarrow_s$, which considers a single statement. In Figure 24 we give the rules for single statements and in Figure 25 the ones for sequences of statements. Finally, in Figure 26 we present the rules for binding and activating method calls. In the following we comment the most significants rules.

**Figure 24** Reduction rules for statements.

Rule (ASGN-FIELD) checks the allowance for action **use** in the accountable entity $\eta'$, then it validates for the entity of the current object $\eta$, the allowance for actions

{**store**, **collect**} for collecting and long-term storing of personal data. Rule (New-Obj) creates a new object with a unique identifier $o''$ and entity $\eta''$ whose fields are initialised to the arguments of the constructor, a reference to $o''$ is bound to $x$ in $\sigma$. The function $fields(C)$ returns the declared fields in the class and the function $\neg personalData(\overline{val})$ returns $true$ if $\overline{val}$ do not contain personal data values, execution will not proceed otherwise; therefore, the rule does not include checks with the $comply$ predicate. Rule (Async-call), a new message is generated. The message contains the caller $o$ and its associated entity $\eta$, the callee $o'$, the arguments of the call $\overline{val}$ and a fresh future $f'$, which is also added to the configuration. The asynchronous call will transfer the rights according to $\eta'$ if the call is made via ! otherwise the call will be served with the rights of the entity of the callee $\eta'''$. Compliance checks the allowance for actions {**use**, **transfer**} in the accountable entity $\eta'$, and according to tags $\oplus_{i \in I} t_i$. Rule (Read-Fut) deferences a future of the form $f'(_t v)$. Note that if the future lacks a value, it is of the form $f$ and the object is blocked until the value in the future is available. Since the future value is dereferenced and assigned to local variable $x$ the corresponding comply case checks the allowance for actions **use** and **collect** in the accountable entity, according to tag $t$. For the Rule (Return), we have the same check that we do for Rule (Async-call), since the statement is also transferring data to the associated future of the method call. Both Rules (Return) and (End-Mtd) replace the object's active process with $idle$.

In Figure 26, in Rule (Bind-Mtd) the process resulting from the binding of the method $m$ on object $o$ with actual parameters $\overline{val}$, associated entity $\eta'$ and future $f$ is put in the queue of $o$. The local variables $\overline{y}$ are initialised with to default values of the types $\overline{A'}$, returned by the function $\overline{def(A')}$. Compliance checks the allowance for action {**use**, **collect**} in the accountable entity according to tag $t$. Note that reduction could get stuck also if the class of the callee does not have a method $m$ with the right number of parameters. Rule (Activate) schedules for execution an enabled process (if possible) from the process queue $q$ of an object $o(\varphi, idle, q, \eta)$.

<div align="center">

(Seq1)

$$\frac{o(\varphi, \{\sigma \mid s\}^f_{\eta'}, ...)^C \; CN \to_s o(\varphi', \{\sigma' \mid s'\}^f_{\eta'}, ...)^C \; CN' \quad s' \neq \textbf{unit}}{o(\varphi, \{\sigma \mid s; sts\}^f_{\eta'}, ...)^C \; CN \to o(\varphi', \{\sigma' \mid s'; sts\}^f_{\eta'}, ...)^C \; CN'}$$

(Seq2)

$$\frac{o(\varphi, \{\sigma \mid s\}^f_{\eta'}, ...)^C \; CN \to_s o(\varphi', \{\sigma' \mid \textbf{unit}\}^f_{\eta'}, ...)^C \; CN'}{o(\varphi, \{\sigma \mid s; sts\}^f_{\eta'}, q, \eta)^C \; CN \to o(\varphi', \{\sigma' \mid sts\}^f_{\eta'}, ...)^C \; CN'}$$

(End-Mtd)

$$o(\varphi, \{\sigma \mid \textbf{unit}\}^f_{\eta'}, ...)^C \; CN \to o(\varphi, idle, q, \eta)^C \; CN$$

(Return)

$$\frac{comply_T(\Sigma, \eta', t) \quad e \mid \langle \varphi, \sigma \rangle \Downarrow_t v}{o(\varphi, \{\sigma \mid \textbf{return}\, e\}^f_{\eta'}, ...)^C \; cn \; f \; \Sigma \to o(\varphi, idle, q, \eta)^C \; cn \; f(_t v) \; \Sigma}$$

</div>

■ **Figure 25**    Reduction rules for sequences of statements.

$$\frac{I = \{i \mid val_i =_{t_i} v_i\} \quad \bigwedge_{i \in I} comply_T(\Sigma, \eta', t_i) \quad B \; m(\overline{A}\,\overline{x})\{\; \overline{A'}\,\overline{y}; \; sts\} \in C'}{\begin{array}{l} o(\varphi, \gamma, q, \eta)^C \; o'(\ldots)^{C'} \; m(o, o', \overline{val}, f, \eta') \; \Sigma \\ \rightarrow o(\varphi, \gamma, q \; \{\overline{x} \mapsto \overline{val} \; \overline{y} \mapsto \overline{def(A')} \mid sts\}^f_{\eta'}, \eta)^C \; o'(\ldots)^{C'} \; \Sigma \end{array}} \text{(\textsc{Bind-Mtd})}$$

$$\frac{\gamma = scheduler(q)}{o(\varphi, idle, q, \eta)^C \rightarrow o(\varphi, \gamma, (q \setminus \gamma), \eta)^C} \text{(\textsc{Activate})}$$

■ **Figure 26** Binding and activation of a process in an object.

## B  Missing Rules of the Type System

In this section, we present the missing type rules for simple expressions, some of the type rules for the single statements and sequence of statements missing from type rules from the main paper. The typing rules for expressions of Figure 27 are

$$\frac{}{\Gamma; \Delta \vdash v : TyOf(v)} \text{(\textsc{t-val})}$$

$$\frac{}{\Gamma; \Delta, [\textbf{this.}]x : RT \vdash [\textbf{this.}]x : RT} \text{(\textsc{t-acc1})} \quad is\text{-}init(RT) \quad \frac{}{\Gamma, [\textbf{this.}]x : DT; \Delta \vdash [\textbf{this.}]x : DT} \text{(\textsc{t-acc2})}$$

$$\frac{}{\Gamma; \Delta, \textbf{this} : C \vdash \textbf{this} : C} \text{(\textsc{t-this})} \quad \frac{\Gamma; \Delta \vdash e_i : T_i \; i \in \{1, .., n\}}{\Gamma; \Delta \vdash (e_1, .. e_n) : T_1 * \cdots * T_n} \text{(\textsc{t-tuple})} \quad \frac{\Gamma; \Delta \vdash e : T_1 * \cdots * T_n \; i \in \{1, .., n\}}{\Gamma; \Delta \vdash e.i : T_i} \text{(\textsc{t-select})}$$

■ **Figure 27** Type rules for simple expressions. Rules (\textsc{t-e-tag}) and (\textsc{t-op}) are in Section 3.

standard. Note that in Rule (\textsc{t-acc1}) we require that the user or or object referred to be initialised, so it is either a parameter or a local variable which has be assigned an initial value.

The typing rules for the statements missing from Section 3 are presented in Figure 28. Rule (\textsc{t-skip}) and Rule (\textsc{t-unit}) are obvious. Rule (\textsc{t-user}) modifies the type of the variable to which the new user is assigned by defining its initial consent that specifies the empty policy. Note that we can only assign to a user variable which is not initialized yet. Assignment to local variables, Rule (\textsc{t-asgn-local}), requires that the type of the expression to be assigned is compatible with the one of the variable. Constraints are generated using the constraint generation case for $Cnstr(\textsc{t-asgn-local}, \Lambda, \Delta)$. The type of the assigned variable, in the resulting environment is modified to reflect the new tag annotation.

The typing of the conditional statement, Rule (\textsc{t-cond}), requires the type of the expression to be a *Bool*. For the typing of the two branches we require the objects/users and future environment produced by the two branches to be the same, also the changes to the consents of users, if any, should be the same. For the environment of data, to get the environment for typing the rest of the statements, for each variable we combine the tag annotations of the two environments.

$$(\text{T-SKIP})$$
$$\overline{\Gamma; \Delta \vdash \textbf{skip} \triangleright \Gamma; \Delta \mid \emptyset}$$

$$(\text{T-UNIT})$$
$$\overline{\Gamma; \Delta \vdash \textbf{unit} \triangleright \Gamma; \Delta \mid \emptyset}$$

$$(\text{T-NEW-USER})$$
$$\overline{\Gamma; \Delta, x : \text{U} \vdash x := \textbf{new user} \triangleright \Gamma; \Delta, x : \text{U}\langle\emptyset\rangle \mid \emptyset}$$

$$(\text{T-ASGN-LOCAL})$$
$$\frac{\Gamma; \Delta \vdash \alpha : T^{\Lambda'}}{\Gamma, x : T^{\Lambda}; \Delta \vdash x := \alpha \triangleright \Gamma, x : T^{\Lambda[\Lambda']}; \Delta \mid Cnstr_T(\Lambda[\Lambda'], \Delta)} \qquad \Lambda' = \epsilon \; \lor \; \Lambda \neq \epsilon$$

$$(\text{T-COND})$$
$$\frac{\Gamma; \Delta \vdash e : Bool^{\Lambda} \quad \Gamma; \Delta \vdash s_1 \triangleright \Gamma_1; \Delta' \mid \boldsymbol{Cn}_1 \quad \Gamma; \Delta \vdash s_2 \triangleright \Gamma_2; \Delta' \mid \boldsymbol{Cn}_2}{\Gamma; \Delta \vdash \textbf{if}\, e \, s_1 \, \textbf{else}\, s_2 \triangleright \Gamma_1 \bullet \Gamma_2; \Delta' \mid \boldsymbol{Cn}_1 \cup \boldsymbol{Cn}_2 \cup Cnstr_U(\Lambda, \Delta)}$$

$$(\text{T-BLOCK})$$
$$\frac{\Gamma; \Delta \vdash sts \triangleright Unit \mid \Gamma'; \Delta' \mid \boldsymbol{Cn}}{\Gamma; \Delta \vdash \{sts\} \triangleright Unit; \Delta' \mid \boldsymbol{Cn}}$$

$$(\text{T-END})$$
$$\overline{\Gamma; \Delta \vdash \textbf{unit} \triangleright Unit \mid \Gamma; \Delta \mid \emptyset}$$

■ **Figure 28** Type rules for single statements and sequences of statements missing from Section 3 .

The composition of environment $\Gamma_1 \bullet \Gamma_2$ is defined provided that the environments are compatible, Proposition 1 and it is defined by

$$(\Gamma_1 \bullet \Gamma_2)(x) \doteq T^{\Lambda_1 \bullet \Lambda_2} \qquad \text{if } \Gamma_i(x) \doteq T^{\Lambda_i} \text{ for } i \in \{1, 2\}$$

It is easy to prove that

**Lemma 1.** $\Gamma_1 \bullet \Gamma_2 \sqsubseteq \Gamma_i$ and $\Gamma_1 \bullet \Gamma_2$ is compatible with $\Gamma_i$ for $i = 1, 2$.

Note that Proposition 1 ensures that the environments produced by the typing of the two branches of the conditional statement are compatible.

Moreover, in addition to the constraints of the two branches and the one of the continuation, we generate the constraints with case $Cnstr(\text{T-COND}, \Lambda, \Delta)$.

For a block to be a correct statement it must be a correct sequence of statements of type $Unit$, i.e., ending with **unit**, Rule (T-BLOCK). Rule (T-END) does not change the typing environments and does not produce constraints.

## C  Proof of Soundness of the Type System for the Typed Operational Semantics

In this section we give the definitions and results necessary to prove Theorem 1.

We extend the type judgements to the runtime configurations. To do so, we need informations on the objects and users defined in the configuration, which is given by the following functions:

- $Os(CN) = \{(o, C, \eta) \mid o(\ldots, \eta)^C \in CN\}$
- $Us(CN) = \{u \mid u \in dom(\Sigma) \; \land \; \Sigma \in CN\}$
- $Fs(CN) = \{f \mid f \in cn\} \cup \{f(_t v) \in CN\}$

We start by defining, in Figure 29, when a *value has a given type* in a configuration *CN* using the above information. For data types we consider both language types and

$$\frac{\substack{TyOf(v)=T \\ (t=\langle \widehat{u}, \widehat{p}\rangle \Longrightarrow \widehat{u} \subseteq Us(CN))}}{CN \models_t v : T^\#} \qquad \frac{TyOf(v)=T}{CN \models v : T} \qquad \frac{}{CN \models u : UT}$$

$$\frac{CT = C \ \lor \ CT = C\langle \eta \rangle \ \lor \ CT = C\langle\_\rangle}{CN \models o : CT}$$

$$\frac{TyOf(v) = T \quad InstAnn(\Lambda, \langle \varphi_0, \sigma_0 \rangle) \sqsubseteq t \quad (t = \langle \widehat{u}, \widehat{p}\rangle \Longrightarrow \widehat{u} \subseteq Us(CN))}{CN \models_{\langle \varphi_0, \sigma_0 \rangle \ t} v : T^\Lambda}$$

■ **Figure 29** Well-typed values in runtime configurations.

extended (tag annotated) types. Only untagged values have a non-personal type $T$, provided that the type of the value is $T$, whereas tagged values may have a personal type provided that their type is right and the tag contains only users defined in $CN$. For extended data types, the tag annotation is instantiated in the runtime object by providing the values of the fields, the parameters and local variables and the resulting tag must be more restrictive than the tag of the data. So the tag annotation allows less actions than the tag of the data, by requiring more users to give their consent for less purposes. A user has type $UT$ if it is defined and an object has a type $CT$ if $CT$ is compatible with its associated class in $CN$.

Configurations contains messages, futures, and objects that may be idle or running in a process. In Figure 30 we first define the judgment asserting that a *message in transit is well-typed* when there is a method with the specified name in the class of the receiver, the corresponding method is well-typed and the type of the actual parameters agree with the type of the formal parameters. Note that the types of the formal parameters are user level types so we do not require the judgment (∗). To check that objects (either idle or running) are well-typed, we have to have that their queue of processes contains the information produced by applications of Rule (Bind-Mtd). This is expressed by the judgment $CN \models q$ **OK in** $C$ which ensures that all the elements of the queue come from well-typed methods of the class $C$, so the process in addition to containing the right body has an environment $\sigma$, associating variables with values that agree with the type of parameters and local variables of the method in the judgment $CN; \Gamma; \Delta \models_{\langle \varphi_0, \sigma_0 \rangle} \langle \varphi, \sigma \rangle$ **OK**. *Idle objects are well-typed* if they have a well-typed queue, and have fields that agree with their declared type. The judgment for *well-typed running objects*, $CN; \Gamma; \Delta \models_{\langle \varphi_0, \sigma_0 \rangle} o(\varphi, \{\sigma \mid ...\}^f_{\eta'}, ...)^C$ **OK**, requires that the object, in addition to have a well-typed queue, also have attributes, parameters and local variables whose type and annotation be in accord with the type environments $\Gamma$ and $\Delta$. Checking the *conformance of the fields and local variables* of an object to to the type environments $\Gamma$ and $\Delta$ is done by the judgment $CN; \Gamma; \Delta \models_{\langle \varphi_0, \sigma_0 \rangle} \langle \varphi, \sigma \rangle$ **OK**. This judgment depends on the values of attributes, parameters and local variables of the method. The information is needed, in the judgement $CN \models_{\langle \varphi_0, \sigma_0 \rangle \ t} v : T^\Lambda$, to instantiate the tags in $\Gamma$, since, looking at the definition of initial type environment of Figure 16, we

$$\frac{(o,C,\eta)\in Os(CN)\quad (o',C',\eta')\in Os(CN)\quad \exists(o'',C'',\eta'')\in Os(CN)\quad f\in Fs(CN)}{B{:}\Lambda\; m(\overline{A\,x})\cdots\; \textbf{OK in } C'\quad CN\models \overline{val}:\overline{A}}{CN\models m(o,o',\overline{val},f,\eta'')\;\textbf{OK}}$$

$$\frac{\langle\{\sigma|sts\}^f_{\eta'}\rangle,\langle\textbf{\textit{Cn}},\textbf{\textit{Um}}\rangle\rangle_m\in q\implies B{:}\Lambda\; m(\overline{A\,x})\{\;\overline{A'}\;\overline{y};\; sts\},\langle\textbf{\textit{Cn}},\textbf{\textit{Um}}\rangle\;\textbf{OK in } C}{f\in Fs(CN)\quad CN\models\sigma(\overline{x})\,\sigma(\overline{y}):\overline{A}\,\overline{A'}}{CN\models q\;\textbf{OK in } C}$$

$$\frac{\begin{array}{l}[\textbf{this.}]x:T^\Lambda\in\Gamma\implies\sigma(x)[\varphi(x)]=_t\sigma(x):{}_tT^\Lambda\\ [\textbf{this.}]x:CT\in\Delta\;\wedge\;is\text{-}init(CT)\implies CN\models\sigma(x)[\varphi(x)]:CT\\ x:UT\in\Delta\;\wedge\;is\text{-}init(UT)\implies CN\models\sigma(x):UT\\ x:\textbf{Fut}\langle T^\Lambda\rangle\in\Delta\implies\sigma(x)=f\wedge(\,f\in Fs(CN)\vee(\,f({}_tv)\in CN\wedge CN\models_{\langle\varphi_0,\sigma_0\rangle}{}_tv:T^\Lambda\,)\end{array}}{CN;\Gamma;\Delta\models_{\langle\varphi_0,\sigma_0\rangle}\langle\varphi,\sigma\rangle\;\textbf{OK}}$$

$$\frac{fields(C)=\overline{A}\,\overline{x}}{CN\models q\;\textbf{OK in } C\quad CN\models\overline{\varphi(x)}:\overline{A}}{CN\models o(\varphi,idle,q,\eta)^C\;\textbf{OK}}\qquad\frac{f\in Fs(CN)\quad\exists(o',C',\eta')\in Os(CN)}{CN;\Gamma;\Delta\models_{\langle\varphi_0,\sigma_0\rangle}\langle\varphi,\sigma\rangle\;\textbf{OK}\quad CN\models q\;\textbf{OK in } C}{CN;\Gamma;\Delta\models_{\langle\varphi_0,\sigma_0\rangle}o(\varphi,\{\sigma\mid\ldots\}^f_{\eta'},\ldots)^C\;\textbf{OK}}$$

$$\frac{\begin{array}{l}msg\in CN\implies CN\models msg\;\textbf{OK}\; f{:}T^t\in\Theta\implies f\in Fs(CN)\vee(\,f({}_tv)\in CN\;\wedge\;TyOf(v)=T\wedge t\sqsubseteq t')\\ (o,C,\eta)\in Os(CN)\implies(CN\models o(\cdot,idle,\cdot,\eta)^C\;\textbf{OK})\vee(\,\exists\;\Gamma,\Delta,\varphi_0,\sigma_0\;CN;\Gamma;\Delta\models_{\langle\varphi_0,\sigma_0\rangle}o(\ldots,\eta)^C\;\textbf{OK})\\ \Sigma(u)=\chi\implies u\in Us(CN)\quad\text{where } CN=cn\;\Sigma\\ \forall o(\ldots,\langle Um,Uc\rangle,\ldots),o'(\ldots,\langle Um',Uc'\rangle,\ldots)\in cn\quad\neg conflict(\langle Um,Uc\rangle,\langle Um',Uc'\rangle)\end{array}}{\Theta\models CN\;\textbf{OK}}$$

■ **Figure 30**  Well-typed messages, queues, environments, runtime objects and configurations.

can see that the variables present in annotations are the ones associated the the tags of the initial values of parameters and fields. Fields and variables must have the type of their declaration and instantiating a tag, in the initial environment we should get a tag less permissive than the one of the value. This will allow us to prove that checking that the compliance of constraints is satisfied at the beginning of the method ensures that compliance will hold in the runtime configuration. The same check is done for variables with future type for which there is a value available. For reference variables we require that they are defined and have type compatible with their declaration in $\Delta$. Finally we define when a *configuration is well-typed,* $\Theta\models CN$ **OK**. To give this judgment we need an environment, $\Theta=\overline{f{:}T^t}$, associating future names to tagged types. Futures are created during execution when we encounter an asynchronous call. In $\Theta$ we record the tagged type of the expected result of the call. For a configuration to be well-typed all the messages must be well-typed and all the futures $f$ in $\Theta$ must be defined in $CN$ and if they contain a personal data its tag must be less less restrictive than the one of the future in $\Theta$. This is because this value will be used in the calling environment according to the consent specified in $\Theta$. All the objects should be well-typed and all the users in the consent environment must be defined in $CN$. Moreover, there should not be two running objects with conflicting sets of modified and checked users.

Subject reduction and progress for expressions is needed for subject reduction for statements.

**Lemma 2** (Subject reduction and progress for expressions). *Let $\Gamma; \Delta \vdash \alpha : ET$ and $CN; \Gamma; \Delta \models_{\langle \varphi, \sigma_0 \rangle} o(\varphi, \{\sigma \mid ...\}^f_{\eta'}, ...)^C$ **OK**. Then $\alpha \mid \langle \varphi, \sigma \rangle \Downarrow val$ and*

- *if $ET = T^\Lambda$ for some $T$ and $\Lambda$, then $CN \models_{\langle \varphi, \sigma_0 \rangle} val : T^\Lambda$,*
- *if $ET = RT$ for some $RT$, then $CN \models val : RT$.*

*Proof.* By induction on $\alpha$. □

We define the relation between environments, $\Gamma; \Delta \rhd \Gamma'; \Delta'$, that shows the evolution of the type environments of a well typed program. For data environments tags of the annotations of variables may change, but personal data cannot be assigned to variables that can be used in an unrestricted way. For reference environments local object and user variables may be initialized and users may add and/or remove consent and the type of the data associated to future variables.

**Definition 5.** *Define $\Gamma; \Delta \rhd \Gamma'; \Delta'$ as $\Gamma \rhd \Gamma'$ and $\Delta \rhd \Delta'$, where*

- *$\Gamma \rhd \Gamma'$ if $\Gamma$ and $\Gamma'$ are compatible and*
- *$\Delta \rhd \Delta'$, if $vr : RT \in \Delta$ ($vr : FT \in \Delta$) implies that $vr : RT' \in \Delta'$ ($x : FT' \in \Delta'$) for some $RT'$ ($FT'$) and either $RT = RT'$ ($FT = FT'$) or*
  1. *$RT = C$ implies $RT' = C\langle \eta \rangle$*
  2. *$RT = \mathsf{U}$ implies $RT' = \mathsf{U}\langle R \rangle$ for some $R$*
  3. *$RT = \mathsf{U}\langle R \rangle$ implies $RT' = \mathsf{U}\langle R' \rangle$ such that $R$ is a prefix of $R'$*
  4. *$FT = \mathbf{Fut}\langle T^\Lambda \rangle$ implies $FT' = \mathbf{Fut}\langle T^{\Lambda'} \rangle$ with $\Lambda' = \epsilon \ \lor \Lambda \neq \epsilon$*

It is easy to show that $\Gamma_1; \Delta_1 \rhd \Gamma_2; \Delta_2$ and $\Gamma_2; \Delta_2 \rhd \Gamma_3; \Delta_3$ implies $\Gamma_1; \Delta_1 \rhd \Gamma_3; \Delta_3$ (the relation is transitive).

We define also a subsumption relation on data environment saying that the tag annotations of the types of variables are more/less restrictive.

**Definition 6.** *We say that $\Gamma \sqsubseteq \Gamma'$ if $[\mathbf{this}.]x : T^\Lambda \in \Gamma$ implies $[\mathbf{this}.]x : T^{\Lambda'} \in \Gamma$ with $\Lambda \sqsubseteq \Lambda'$.*

Subject reduction for statements, Lemma 3, and sequences of statements, Lemma 4, are proved, as usual, by induction on the rules of the reduction relations $\to$ and $\to_s$.

**Lemma 3** (Subject reduction for statements). *Consider an object executing method. Let $\langle \varphi_0, \sigma_0 \rangle$ and $\Sigma_0$ be the attributes, the local environment and the consent at the beginning of the execution of the current method, and let $\langle Um, Uc \rangle$ be the modified and checked users of the method.*
*Let $CN = o(\varphi, \{\sigma \mid s\}^f_{\eta'}, \langle Um, Uc \rangle, q, \eta)^C \ cn \ \Sigma$ with*
*(1) if $u \in Um \cup Uc$ then $\Sigma(u) = \Sigma_0[\Delta](u)$ and*
*(2) $\Theta \models CN$ **OK** for some $\Theta$ and*
*(3) $CN; \Gamma; \Delta \models_{\langle \varphi_0, \sigma_0 \rangle} o(\varphi, \{\sigma \mid s\}^f_{\eta'}, q, \eta)^C$ **OK** and*
*(4) $\Gamma; \Delta \vdash s \rhd \Gamma_1; \Delta_1 \mid \boldsymbol{Cn_1}$.*

If $o(\varphi, \{\sigma \mid s\}_{\eta'}^{f}, \ldots)^C$ cn $\Sigma \rightarrow_s CN' = o(\varphi', \{\sigma' \mid s'\}_{\eta'}^{f}, \ldots)^C$ cn' $\Sigma'$, then there are $\Gamma'$, $\Gamma''$, $\Delta'$ and **Cn** such that

(i) $\Gamma; \Delta \triangleright \Gamma'; \Delta'$ and $\Gamma_1 \sqsubseteq \Gamma''$ and **Cn** $\subseteq$ **Cn**$_1$ and

(ii) $\Gamma'; \Delta' \vdash s' \triangleright \Gamma''; \Delta_1 \mid$ **Cn** and

(iii) $\Theta' \models CN'$ **OK** for some $\Theta'$ such that $\Theta' \setminus \Theta = \Theta$ and

(iv) $\Theta'; CN'; \Gamma'; \Delta' \models_{\langle \varphi_0, \sigma_0 \rangle} o(\varphi', \{\sigma' \mid s'\}_{\eta'}^{f}, \ldots)^C$ **OK**

(v) if $u \in Um$ then $\Sigma'(u) = \Sigma_0[\Delta'](u)$ otherwise $\Sigma'(u) = \Sigma(u)$.

*Proof.* By cases on the rules of Figure 24 using Lemma 2. □

**Lemma 4** (Subject reduction for sequences of statements). *Consider an object executing a method and let $\langle \varphi_0, \sigma_0 \rangle$ and $\Sigma_0$ be the attributes, the local environment and the consent at the beginning of the execution of the current method, and let $\langle Um, Uc \rangle$ be the modified and checked users of the method.*
Let $CN = o(\varphi, \{\sigma \mid sts\}_{\eta'}^{f}, \langle Um, Uc \rangle, q, \eta)^C$ cn $\Sigma$ with

(1) *if* $u \in Um \cup Uc$ *then* $\Sigma(u) = \Sigma_0[\Delta](u)$ *and*

(2) $\Theta \models CN$ **OK** *for some* $\Theta$ *and*

(3) $CN; \Gamma; \Delta \models_{\langle \varphi_0, \sigma_0 \rangle} o(\varphi, \{\sigma \mid sts\}_{\eta'}^{f}, q, \eta)^C$ **OK** *and*

(4) $\Gamma; \Delta \vdash sts \triangleright DT \mid \Gamma_1; \Delta_1 \mid$ **Cn**$_1$.

*If* $o(\varphi, \{\sigma \mid sts\}_{\eta'}^{f}, \ldots)^C$ *cn* $\Sigma \rightarrow CN' = o(\varphi', \{\sigma' \mid sts'\}_{\eta'}^{f}, \ldots)^C$ *cn'* $\Sigma'$, *where* $sts \neq$ **unit** *and* $sts \neq$ **return** $\alpha$, *then then there are* $\Gamma'$, $\Gamma''$, $\Delta'$ *and* **Cn** *such that*

(i) $\Gamma; \Delta \triangleright \Gamma'; \Delta'$ *and* $\Gamma_1 \sqsubseteq \Gamma''$ *and* **Cn** $\subseteq$ **Cn**$_1$ *and*

(ii) $\Gamma'; \Delta' \vdash sts' \triangleright DT \mid \Gamma''; \Delta_1 \mid$ **Cn** *and*

(iii) $\Theta' \models CN'$ **OK** *for some* $\Theta'$ *such that* $\Theta' \setminus \Theta = \Theta$ *and*

(iv) $CN'; \Gamma'; \Delta' \models_{\langle \varphi_0, \sigma_0 \rangle} o(\varphi', \{\sigma' \mid sts'\}_{\eta'}^{f}, \ldots)^C$ **OK** *and*

(v) *if* $u \in Um$ *then* $\Sigma'(u) = \Sigma_0[\Delta'](u)$ *otherwise* $\Sigma'(u) = \Sigma(u)$.

*Proof.* By cases on the rules of Figure 25 using Lemma 3. □

**Lemma 5** (Subject Reduction for End of Processes). *Consider an object executing a method and let $\langle \varphi_0, \sigma_0 \rangle$ and $\Sigma_0$ be the attributes, the local environment and the consent at the beginning of the execution of the current method, and let $\langle Um, Uc \rangle$ be the modified and checked users of the method.*
Let $CN = o(\varphi, \{\sigma \mid sts\}_{\eta'}^{f}, \langle Um, Uc \rangle, q, \eta)^C$ cn $\Sigma$ with

(1) *if* $u \in Um \cup Uc$ *then* $\Sigma(u) = \Sigma_0[\Delta](u)$ *and*

(2) $\Theta \models CN$ **OK** *for some* $\Theta$ *and*

(3) $CN; \Gamma; \Delta \models_{\langle \varphi_0, \sigma_0 \rangle} o(\varphi, \{\sigma \mid sts\}_{\eta'}^{f}, q, \eta)^C$ **OK** *where* $sts =$ **return** $e$ *or* $sts =$ **unit** *and*

(4) $\Gamma; \Delta \vdash sts \triangleright T^{\Lambda} \mid \Gamma_1; \Delta_1 \mid$ **Cn**$_1$.

*If* $o(\varphi, \{\sigma \mid sts\}_{\eta'}^{f}, \langle Um, Uc \rangle, q, \eta)^C$ *cn* $\Sigma \rightarrow CN' = o(\varphi, idle, q, \eta)^C$ *cn'* $\Sigma$ *then* $\Theta \models CN'$ **OK**.
*If* $sts =$ **return** $e$ *and* $e \mid \langle \varphi, \sigma \rangle \Downarrow_t v$, *then we also have* $CN' \models_{\langle \varphi_0, \sigma_0 \rangle} {}_t v : T^{\Lambda}$.

*Proof.* By cases on the rules of Figure 25. □

**Lemma 6** (Subject Reduction for Rules (Bind-Mtd) and (Activate)).

**1.** Let $CN = (\varphi, \gamma, q, \eta)^C \ m(o, o', \overline{val}, f, \eta') \ cn \ \Sigma$ _and_ $\Theta \models CN$ **OK** _for some_ $\Theta$. _If_
$$CN \rightarrow CN' = o(\varphi, \gamma, q \ \langle \{\overline{x} \mapsto \overline{val} \ \overline{y} \mapsto \overline{def(A')} \mid sts\}^f_{\eta'}, \langle \boldsymbol{Cn}, \boldsymbol{Um} \rangle\rangle_m, \eta)^C \ cn \ \Sigma$$
_with Rule_ (BIND-MTD), _then_ $\Theta \models CN'$ **OK**.

**2.** Let $CN = o(\varphi, idle, p \ q, \eta)^C \ cn \ \Sigma$ _where_ $p = \langle\{\sigma \mid sts\}^f_{\eta'}, \langle \boldsymbol{Cn}, \boldsymbol{Um} \rangle\rangle_m$ _and_ $\Theta \models CN$ **OK** _for some_ $\Theta$. _If_ $CN \rightarrow CN' = o(\varphi, \{\sigma \mid sts\}^f_{\eta'}, \langle Um, Uc \rangle, (q \setminus p), \eta)^C \ cn \ \Sigma$ _with Rule_ (ACTIVATE), _then_

  (i) $\Theta \models CN'$ **OK** _and_

  (ii) $CN'; \Gamma_{lc}, \Gamma_{p\&f}; \Delta_{lc}, \Delta_{p\&f} \models_{\langle \varphi, \sigma \rangle} o(\varphi, \{\sigma \mid \cdots\}^f_{\eta'}, (q \setminus p), \eta)^C$ **OK** _and_

  (iii) $\Gamma_{lc} \ \Gamma_{p\&f}; \Delta_{lc} \ \Delta_{p\&f}, \textbf{this} : C \vdash sts \rhd T^{\Lambda'} \mid \Gamma; \Delta \mid \boldsymbol{Cn}$ _and_

  (iv) $\forall \pi \in InstCnstr(\boldsymbol{Cn}, \langle \varphi, \sigma \rangle, \eta', \eta, \Sigma) \quad \pi$ _holds._

_Proof._ By cases on the rules of Figure 19. □

Note that Point iv of the previous lemma holds for the consent environment at the beginning of the execution of the method.

It is easy to show that the initial configuration is well-typed.

**Lemma 7** (Well-typed initial configuration). _Let_ $CN$ _be_
$$main(\emptyset, \{\overline{x} \mapsto \overline{def(A)}, \textbf{this} \mapsto main \mid sts\}_\bot, \langle \emptyset, \emptyset \rangle, \bot)^{Main} \ \emptyset$$
_where_

- $\{\overline{A} \ \overline{x}; sts\}$ _is the main block of PR and_
- $\Gamma_{lc}; \Delta_{lc} \vdash sts \rhd Unit \mid \Gamma; \Delta \mid \emptyset$ _and where_ $(\Gamma_{lc}, \Delta_{lc}) = TE_{lc}(\overline{A} \ \overline{x})$.

_Then_ $\emptyset \models CN$ **OK**.

**Lemma 8** (Monotonicity of compliance). _If_ $act \in \mathscr{A}(\Sigma, \eta', t')$ _and_ $t' \sqsubseteq t$, _then_ $act \in \mathscr{A}(\Sigma, \eta', t)$

**Lemma 9** (Well-typed reachable configurations). _Let_ $CN = cn \ \Sigma$ _be a reachable configurations. Then_

  (i) $\Theta \models CN$ **OK** _for some_ $\Theta$

_and for all_ $o(\varphi, \{\sigma \mid sts\}^f_{\eta'}, \langle Um, Uc \rangle, q, \eta)^C \in CN$, _let_ $\langle \varphi_0, \sigma_0 \rangle$ _and_ $\Sigma_0$ _be the attributes, the local environment and the consent at the beginning of the execution of the current method and let_ $\langle Um, Uc \rangle$ _be the modified and checked users of the method, we have_

  (ii) _if_ $u \in Um \cup Uc$ _then_ $\Sigma(u) = \Sigma_0[\Delta](u)$ _and_

  (iii) $CN; \Gamma; \Delta \models_{\langle \varphi_0, \sigma_0 \rangle} o(\varphi, \{\sigma \mid sts\}^f_{\eta'}, q, \eta)^C$ **OK** _and_

  (iv) $\Gamma; \Delta \vdash sts \rhd DT \mid \Gamma_1; \Delta_1 \mid \boldsymbol{Cn}$ _and_

  (v) $\forall \pi \in InstCnstr(\boldsymbol{Cn}, \langle \varphi_0, \sigma_0 \rangle, \eta', \eta, \Sigma_0) \quad \pi$ _holds._

_Proof._ By induction on the number of reduction steps, $n$, from the initial configuration. For $n = 0$, from Lemma 7.

Let $CN'$ be obtained by $n$ reduction steps with $n > 0$ and let the last reduction be $CN \rightarrow CN'$ where $CN = cn \ \Sigma$. By inductive hypothesis we have that

(1) $\Theta \models CN$ **OK** _for some_ $\Theta$

and for all $o(\varphi, \{\sigma \mid sts\}^f_{\eta'}, \langle Um, Uc \rangle, q, \eta)^C \in cn$, let $\langle \varphi_0, \sigma_0 \rangle$ and $\Sigma_0$ be the attributes, the local environment and the consent at the beginning of the execution of the current method and let $\langle Um, Uc \rangle$ be the modified and checked users of the method, we have

(2) if $u \in Um \cup Uc$ then $\Sigma(u) = \Sigma_0[\Delta](u)$ and

(3) $CN; \Gamma; \Delta \models_{\langle \varphi_0, \sigma_0 \rangle} o(\varphi, \{\sigma \mid sts\}^f_{\eta'}, q, \eta)^C$ **OK** and

(4) $\Gamma; \Delta \vdash sts \triangleright DT \mid \Gamma_1; \Delta_1 \mid \boldsymbol{Cn}$ and

(5) $\forall \pi \in InstCnstr(\boldsymbol{Cn}, \langle \varphi_0, \sigma_0 \rangle, \eta', \eta, \Sigma_0)$   $\pi$ holds.

Consider the reduction $cn\ \Sigma \to CN'$. We have the following cases:

(a) $CN = o(\varphi, \{\sigma \mid sts\}^f_{\eta'}, ...)^C\ cn''\ \Sigma$ and $CN' = o(\varphi', \{\sigma' \mid sts'\}^f_{\eta'}, ...)^C\ cn'\ \Sigma'$

(b) $CN = o(\varphi, \{\sigma \mid sts\}^f_{\eta'}, \langle Um, Uc \rangle, q, \eta)^C\ cn''\ \Sigma$ and $CN' = o(\varphi, idle, q, \eta)^C\ cn'\ \Sigma$

(c) $CN = (\varphi, \gamma, q, \eta)^C\ m(o, o', \overline{val}, f, \eta')\ cn''\ \Sigma$ and
$CN' = o(\varphi, \gamma, q\ \langle\{\overline{x} \mapsto \overline{val}\ \overline{y} \mapsto \overline{def(A')} \mid sts\}^f_{\eta'}, \langle \boldsymbol{Cn}, \boldsymbol{Um}\rangle\rangle_m, \eta)^C\ cn''\ \Sigma$

(d) $CN = o(\varphi, idle, p\ q, \eta)^C\ cn''\ \Sigma$ where $p = \langle\{\sigma|sts\}^f_{\eta'}, \langle \boldsymbol{Cn}, \boldsymbol{Um}\rangle\rangle_m$ and
$CN' = o(\varphi, \{\sigma|sts\}^f_{\eta'}, \langle Um, Uc \rangle, (q \setminus p), \eta)^C\ cn''\ \Sigma$.

From Point 1 we derive that if $u \in Um$ of a running object $o(..., \langle Um, Uc \rangle, ...) \in cn$, then $u \notin Um' \cup Uc'$ for all other running objects $o'(..., \langle Um', Uc' \rangle, ...) \in cn$. Notice that $\Sigma \neq \Sigma'$ only if the reduction was the one of Point a, and in this case after the reduction $\Sigma'(u) = \Sigma[\Delta'](u)$ where $\Delta'$ is the reference environment typing $sts'$. Therefore, Point 2 holds for all running objects in $CN'$. Moreover Point 5 holds for all objects $o'(..., \langle Um', Uc' \rangle, ...) \in CN'$, and it holds for $o$ from the fact that the process was generated by Rule (Activate). Therefore Points ii and v hold for all running objects.

For Case (a) we derive Points (i) and (iii) (iv) from Lemma 4.

For Case (b) we derive Points (i) and (iii) (iv) from subject reduction for end of processes.

For Case (c) and (d) we derive Points (i) and (iii) (iv) from Lemma 6.   □

From the previous lemmas we can prove Theorem 1.

**Theorem 1** [Progress for running objects] *Let* $o(\varphi, \{\sigma \mid sts\}^f_{\eta'}, \langle Um, Uc \rangle, q, \eta)^C\ CN$ *be reachable from PR, then for some* $CN'$

$$o(\varphi, \{\sigma \mid sts\}^f_{\eta'}, \langle Um, Uc \rangle, q, \eta)^C\ CN \to CN'$$

*or* $sts = y := x.\boldsymbol{get}$ *and* $\sigma(x) = f'$ *and* $f' \in CN$.

*Proof.* Let $CN_0 = o(\varphi, \{\sigma \mid sts\}^f_{\eta'}, \langle Um, Uc \rangle, q, \eta)^C\ cn\ \Sigma$ be reachable from $PR$. By Lemma 9 we have that

(i) $\Theta \models CN_0$ **OK** for some $\Theta$

and for all $o(\varphi, \{\sigma \mid sts\}^f_{\eta'}, \langle Um, Uc \rangle, q, \eta)^C \in cn$, let $\langle \varphi_0, \sigma_0 \rangle$ and $\Sigma_0$ be the attributes, the local environment and the consent at the beginning of the execution of the current method and let $\langle Um, Uc \rangle$ be the modified and checked users of the method, we have

(ii) if $u \in Um \cup Uc$ then $\Sigma(u) = \Sigma_0[\Delta](u)$ and

(iii) $CN; \Gamma; \Delta \models_{\langle \varphi_0, \sigma_0 \rangle} o(\varphi, \{\sigma \mid sts\}^f_{\eta'}, q, \eta)^C$ **OK** and

(iv) $\Gamma; \Delta \vdash sts \triangleright DT \mid \Gamma_1; \Delta_1 \mid \boldsymbol{Cn}$ and

(v) $\forall \pi \in InstCnstr(\boldsymbol{Cn}, \langle \varphi_0, \sigma_0 \rangle, \eta', \eta, \Sigma_0)$   $\pi$ holds.

The proof is by cases on the shape of the sequence of statements. We do some significant cases. The others can be proved in a similar way.

**$sts = $ unit** With Rule (END-MTD) $o(\varphi, \{\sigma \mid \mathbf{unit}\}^f_{\eta'}, q, \eta)^C \, CN \rightarrow o(\varphi, idle, q, \eta)^C \, CN$

**$sts = $ return $\alpha$** From (iv) and Rule (T-RETURN)
$$\Gamma; \Delta \vdash \mathbf{return} \; \alpha \rhd T^\Lambda \mid \Gamma; \Delta \mid Cnstr_T(\Lambda, \Delta)$$
where $\Gamma; \Delta \vdash \alpha : T^\Lambda$. From Lemma 2 we have $\alpha \mid \langle \varphi, \sigma \rangle \Downarrow val$ and $CN \models_{\langle \varphi_0, \sigma_0 \rangle} val : T^\Lambda$.
From Point (v) and $Cnstr_T(\Lambda, \Delta) \subseteq Cn$ we have
$$\text{for all } \pi \in InstCnstr(Cnstr_T(\Lambda, \Delta), \langle \varphi_0, \sigma_0 \rangle, \eta', \eta, \Sigma_0) \quad \pi \text{ holds}$$
By Definition 2, instantiation of $A_{Cnstr}(act, \Lambda, \Delta)$ produces $act \in \mathscr{A}(\Sigma_0[\Delta], \eta', \langle \widehat{u}, \widehat{p} \rangle)$ where $\langle \widehat{u}, \widehat{p} \rangle = InstAnn(\Lambda, \langle \varphi_0, \sigma_0 \rangle)$. Morever, $\widehat{u} \subseteq Uc$. From Point (ii), $\Sigma(u) = \Sigma_0[\Delta](u)$ for all $u \in \widehat{u}$, and from $CN \models_{\langle \varphi_0, \sigma_0 \rangle} val : T^\Lambda$ we have $InstAnn(\Lambda, \langle \varphi_0, \sigma_0 \rangle) \sqsubseteq t$. Therefore $act \in \mathscr{A}(\Sigma, \eta', t')$ where $t' \sqsubseteq t$ hold. By Lemma 8 we get that $act \in \mathscr{A}(\Sigma, \eta', t)$. Finally from (iii), $cn = cn' \, f$ Therefore, with Rule (RETURN)
$$o(\varphi, \{\sigma \mid \mathbf{return} \, \alpha\}^f_{\eta'}, q, \eta)^C \, cn' \; f \; \Sigma \rightarrow o(\varphi, idle, q, \eta)^C \, cn' \; f(_t v) \; \Sigma$$

**$sts = $ Fut$\langle B \rangle \; x := vr@m'(\overline{\alpha}); sts'$** From (iv) and Rule (T-SEQ) and Rule (T-ASYN-CALL) we get

(a) $\Gamma; \Delta \vdash sts \rhd DT \mid \Gamma_1; \Delta_1 \mid Cn = Cn_1 \cup Cn_2$ and

(b) $\Gamma; \Delta \vdash \mathbf{Fut}\langle B' \rangle \, x := vr@m'(\overline{\alpha}) \rhd \Gamma; \Delta, vr : CT, x : \mathbf{Fut}\langle T^{\Lambda'} \rangle \mid Cn_1$ and

(c) $vr : CT \in \Delta$ and $Cn_1 = \bigcup_{j \in J} Cnstr_T(\Lambda_j, \Delta)$ and

(d) $J = \{j \mid ET_j = T_j^{\Lambda_j} \text{ for some } T_j^{\Lambda_j}\}$ and

(e) $\Gamma; \Delta \vdash \alpha_i : ET_i \quad i \in \{1, .., n\}$.

From (b) and (c) and (iii) we get that $o'(\ldots, \underline{\eta'''})^{C'} \in cn$.
From (e) and Lemma 2 we have $\overline{\alpha} \mid \langle \varphi, \sigma \rangle \Downarrow \overline{val}$ and $CN \models_{\langle \varphi_0, \sigma_0 \rangle} val_i : T^{\Lambda_i}$ for all $i \in \{1, .., n\}$.
In order to apply Rule (ASYNC-CALL),
we have to establish that $\bigwedge_{i \mid val_i = t_i v_i} comply_T(\Sigma, \eta', t_i)$ holds.
As for the previous case, from Point (v) and $Cn_1 = \bigcup_{j \in J} Cnstr_T(\Lambda_j, \Delta) \subseteq Cn$ we have
$$\text{for all } \pi \in InstCnstr(Cn_1, \langle \varphi_0, \sigma_0 \rangle, \eta', \eta, \Sigma_0) \quad \pi \text{ holds}$$
From (d) we have that $J \subseteq I$. From $CN \models_{\langle \varphi_0, \sigma_0 \rangle} val_j : T^{\Lambda_j}$ for all $j \in J$ we can conclude as in the previous case that
$$\bigwedge_{i \mid val_i = t_i v_i} comply_T(\Sigma, \eta', t_i) \text{ holds}$$
Therefore, from Rule (SEQ2) with Rule (ASYNC-CALL) derives that
$$o(\varphi, \{\sigma \mid sts\}^f_{\eta'}, \ldots)^C \; o'(\ldots, \eta''')^{C'} \; cn' \; \Sigma$$
$$\rightarrow_s o(\varphi, \{\sigma[x \mapsto f'] \mid sts'\}^f_{\eta'}, \ldots)^C \; o'(\ldots, \eta''')^{C'} \; m(o', o, \overline{val}, f', \eta'') f'cn' \; \Sigma$$

**$sts = x.\mathbf{addCon}(\rho); sts'$** From Rule (SEQ2) with Rule (ADD-CONSENT) we get
$$o(\varphi, \{\sigma \mid sts\}^f_{\eta'}, \ldots)^C \, cn \; \Sigma \rightarrow o(\varphi, \{\sigma \mid sts'\}^f_{\eta'}, \ldots)^C \, cn \; \Sigma[u \mapsto InsrtPol(\chi, \rho)]$$

<div align="right">□</div>

## About the authors

**Chinmayi Prabhu Baramashetru** Contact her at cpbarama@ifi.
uio.no.
https://orcid.org/0000-0001-5344-0032

**Paola Giannini** Contact her at paola.giannini@uniupo.it.
https://orcid.org/0000-0003-2239-9529

**Silvia Lizeth Tapia Tarifa** Contact her at sltarifa@ifi.uio.no.
https://orcid.org/0000-0001-9948-2748

**Olaf Owe** Contact him at olaf@ifi.uio.no.
https://orcid.org/0000-0003-0976-5678